\documentclass[hidelinks,onefignum,onetabnum]{siamart251216}
\usepackage{amsmath,hyperref}
\usepackage{epstopdf}

\usepackage{lineno,hyperref}
\usepackage{graphicx}
\usepackage{dcolumn}
\usepackage{bm}
\usepackage{epsfig}
\usepackage{booktabs}
\usepackage{subfigure}
\usepackage{graphics}
\usepackage{graphicx}
\usepackage{amssymb}
\usepackage{amsmath}
\usepackage{array}
\usepackage{color}
\usepackage{booktabs}
\usepackage{multirow}
\usepackage{caption}
\usepackage{chngpage}
\usepackage{subfigure}
\usepackage{mathrsfs,gensymb,float}

\usepackage{lipsum}
\usepackage{amsfonts}
\usepackage{graphicx}
\usepackage{epstopdf}
\usepackage{algorithmic}
\ifpdf
  \DeclareGraphicsExtensions{.eps,.pdf,.png,.jpg}
\else
  \DeclareGraphicsExtensions{.eps}
\fi

\newsiamremark{remark}{Remark}
\newsiamremark{hypothesis}{Hypothesis}
\crefname{hypothesis}{Hypothesis}{Hypotheses}
\newsiamthm{claim}{Claim}
\newsiamremark{fact}{Fact}
\crefname{fact}{Fact}{Facts}

\headers{Onsager-based LBM For Three-phase Dielectric  Flows}{X. Liu, X. Qian, F. Xiong and L. Wang}

\title{Onsager-variational-principle-based lattice Boltzmann model for three-phase dielectric fluid flows\thanks{Submitted to the June 1, 2026
\funding{This work was funded by the National Natural Science Foundation of China (Grants No.12371374 and No. 12472297).}}}

\author{Xinyue Liu\thanks{College of Science, National University of Defense Technology, Changsha 410073,  China.}
\and Xu Qian\thanks{College of Science, National University of Defense Technology, Changsha 410073, China
  (\email{qianxu@nudt.edu.cn}).}
\and Fang Xiong\thanks{School of Mathematics and Physics, China University of Geosciences, Wuhan 430074, China.}
\and Lei Wang\footnotemark[4]}

\usepackage{amsopn}

\ifpdf
\hypersetup{
  pdftitle={Onsager-variational-principle-based Lattice Boltzmann Model For Three-phase Dielectric Fluid Flows},
  pdfauthor={X. Liu, X. Qian, F. Xiong and L. Wang}
}
\fi

\begin{document}

\maketitle

\begin{abstract}
Multiphase electrohydrodynamic (EHD) flows play a crucial role in various engineering applications. However, existing numerical studies on three-phase electrohydrodynamic systems predominantly rely on phenomenological models, often neglecting thermodynamic consistency and critical surface charge convection mechanisms. To address these fundamental gaps, this paper proposes a thermodynamically consistent three-phase EHD model derived strictly from the Onsager variational principle. This theoretical framework intrinsically guarantees thermodynamic consistency and accurately captures complex multiphysics interactions without requiring a priori  assumptions. Furthermore, a mesoscopic lattice Boltzmann method  is developed to solve the proposed model, enabling the natural capture of interfacial evolution and charge transport. The accuracy  of the numerical framework are rigorously validated against several benchmark cases, including electroosmotic flow in microchannels, the spreading of a three-phase liquid lens, the equilibrium of static compound droplet, and the deformation of compound droplet under uniform electric field. Using this validated framework, we  investigate  EHD applications, specifically simulating the complex dynamics of double droplet coalescence and separation under electric field, as well as the behavior of droplets subjected to combined EHD and shear flow. Overall, this work provides a robust, thermodynamically  reliable numerical tool for exploring the highly nonlinear behaviors of multiphase EHD systems.
\end{abstract}



\section{Introduction}

Multi-phase electrohydrodynamics (EHD) represents  an interdisciplinary field focused on a pivotal role in a wide range of engineering and scientific applications, such as electrospray technology, electrospinning, microfluidic behavior under electric field, and ink-jet printing \cite{Park:NM2007,Prabhu:NRM2023,Wang:JFM2025}. Consequently, a systematic investigation into the underlying physical mechanisms of multi-phase EHD and a detailed analysis of its evolutionary processes have become essential prerequisites for further exploring and optimizing these applications.

As a significant branch of EHD flow, two-phase EHD phenomena have been extensively investigated through theoretical, experimental, and numerical approaches. The pioneering work of Taylor \cite{Taylor:PR1964} established the classical leaky dielectric model, providing a foundational theoretical framework for understanding the deformation of weakly conducting droplets in uniform electric fields. However, this model neglects the convection of surface charges. Feng et al. \cite{Feng:JFM1996} later demonstrated that when the electric Reynolds number is non-negligible, the convective transport of surface charges significantly influences the interfacial charge distribution and the resulting dynamic behavior. This insight has prompted subsequent studies to universally incorporate surface charge convection mechanisms into EHD models to more accurately reflect physical reality. Building upon these progressively refined theoretical frameworks, researchers have conducted extensive and explorations into the interfacial dynamics and diverse applications of two-phase EHD flows. On the theoretical plane, Saville et al. \cite{Saville:ARF1997} provided a systematic review of electrohydrodynamics, addressing the leaky dielectric model, which established a fundamental framework for the field. Guided by these principles, Liu et al. \cite{Liu:FR2023} examined droplet deformation driven by coupled electric and shear fields, thereby confirming theoretical frameworks. Concurrently, from an engineering perspective, Mhatre et al.\cite{Mhatre:CEP2015} successfully applied electric field technology to the separation of water from emulsified oils. Addressing more complex systems such as colloidal particles, Wang et al. \cite{Wang:JFM2024} investigated the nonlinear EHD flows generated around particles, revealing complex behaviors that remain elusive within classical linear frameworks.

Despite the extensive body of research on two-phase EHD flows, many engineering applications frequently involve more intricate multiphase flows consisting of three or more fluids. For instance, in the design of multiphase microfluidic reactors and the synthesis of advanced functional materials, it is often necessary to manage complex systems characterized by multiple interfaces \cite{Bei:APL2008,Abbasi:SM2019,Huang:SM2019}. Through microfluidic experiments, Bei et al. \cite{Bei:APL2008} demonstrated that this approach can effectively achieve precise centering and controlled steady-state morphologies of compound droplets. However, this steady-state control breaks down under stronger electric field.  Abbasi et al. \cite{Abbasi:SM2019} revealed that strong electric fields trigger multiple interfacial instabilities and various droplet breakup modes in double emulsions. To better align with practical engineering environments, Huang et al. \cite{Huang:SM2019} applied AC electric fields within microchannels, demonstrating precise dynamic control over droplet formation and number. In contrast to two-phase EHD flows, multiphase systems involving three or more fluids feature additional interfaces that trigger highly complex multiphysics interactions, making it challenging to directly extrapolate their dynamic behaviors from classical two-phase theories. Over the past few years, researchers have made  extensive efforts to investigate this topic through experimental,and theoretical approaches \cite{Liu:FR2023,Mhatre:CEP2015,Wang:JFM2024,Bei:APL2008,Abbasi:SM2019,Huang:SM2019,Behjatian:SMP2013}. Nevertheless, theoretical analysis remains feasible only for highly idealized multiphase EHD problems characterized by low Reynolds numbers and simplified interfacial configurations. With the rapid development of computer techology, numerical analysis has become a prevalent tool for investigating flow behavior in multiphase EHD systems, particularly for cases that remain experimentally unexplored or are challenging to study physically due to high costs and extended durations. Regarding numerical methods for multiphase EHD simulations, the literature primarily distinguishes between sharp-interface and diffuse-interface methods. Sharp-interface approaches, such as the volume of fluid (VOF) \cite{Tomar:JCP2007,Das:JFM2021} and level set (LS) \cite{Abbasi:JE2017,Paknemat:POF2012,Naz:POF2023} methods, characterize multiphase systems by explicitly tracking the interface positions. Yet, the intrinsic limitations of VOF and LS methods in calculating surface tension and maintaining mass conservation pose significant challenges for simulating the internal structures of complex multiphase flows. \cite{Worner:MN2012}. In contrast, as a diffuse-interface approach, the phase-field method has garnered widespread attention for its efficacy in handling complex multiphase flows, owing to its ability to describe phase boundaries through a diffuse interface \cite{Santra:IJNMT2021,Yang:JMF2013,Feng:AMM2025}. In recent years, an increasing number of researchers have adopted this method to investigate  EHD problems. Soni et al. \cite{Soni:JE2013} employed the phase-field method to explore the dynamic behavior of compound droplets in a uniform electric field, yielding results in high agreement with experimental observations. Santra et al. \cite{Santra:POF2020} further utilized this method to investigate the   compound droplets  deformation in shear flows, successfully broadening the research focus from isolated electric fields to coupled electro-shear environments. Subsequently, the same research group \cite{Santra:JFM2020} explored these droplets under the dual influence of electric fields and physical confinement, to closely reflect actual experimental conditions in microfluidic devices. However, as the core and continuous phases in these aforementioned studies are identical, the phase-field models employed remained essentially two-phase formulations. Building upon these works, Su et al. \cite{Su:SP2020} implemented a phase-field simulation to address the EHD behavior of true three-phase compound droplets, where the core, shell, and continuous phases all possess distinct physical properties.

The aforementioned phase-field studies have established a solid foundation for understanding multiphase EHD behaviors. Yet, it should be noted that the majority of these works focus on  two-phase EHD problems \cite{Soni:JE2013,Santra:POF2020,Santra:JFM2020}, while research involving three-phase EHD systems \cite{Su:SP2020} remains markedly limited. More critically, the few existing studies on three-phase EHD often employ phenomenological models that simply couple electric and flow fields, leaving the fundamental issue of thermodynamic consistency largely unaddressed. Furthermore, to simplify the governing equations, some researchers still apply the classical leaky dielectric model directly to three-phase EHD problems \cite{Su:SP2020,Liu:PD2024}, which inherently neglects transient charge relaxation and charge convection effects. While this offers computational convenience, previous works \cite{Feng:JFM1996,Luo:POF2020} has demonstrated that such simplifications become inadequate at high electric Reynolds numbers. Recently, Zhang et al. \cite{Zhang:JCP2024} established a thermodynamically consistent three-phase EHD model derived from the principle of energy dissipation. Utilizing the finite difference method, they simulated ternary droplet coalescence and spinodal decomposition, yielding results unattainable in binary systems. However, directly solving such non-linear coupled equations using the finite difference method often leads to massive computational overhead, posing significant challenges for large-scale parallelization and three-dimensional extensions. In light of this, enriching the theoretical framework of three-phase EHD not only requires introducing alternative thermodynamic derivation mechanisms, but on the numerical front, it also urgently demands the development of more efficient algorithms suitable for large-scale computation. To address the physical aspect, this work intends to develop a three-phase EHD physical model using the Onsager variational principle \cite{Eck:IFB2009,Xu:JPCM2015,Xu:PI2017,Chen:JCP2025,Xiao:JCP2025}. Unlike traditional derivations based on energy dissipation \cite{Zhang:JCP2024}, the Onsager principle is rooted in the second law of thermodynamics and a variational framework, without requiring any a priori assumption. Since this methodology mathematically ensures that the resulting governing equations satisfy Onsager's reciprocal relations, it guarantees thermodynamic consistency from its very foundation. Consequently, it has been widely adopted for modeling various complex physical systems. Nevertheless, studies reporting thermodynamically consistent three-phase EHD models based on the Onsager principle are still rare. Furthermore, on the numerical front, departing from conventional macroscopic methods \cite{Soni:JE2013,Santra:POF2020,Santra:JFM2020,Su:SP2020,Zhang:JCP2024}, this paper employs the mesoscopic Lattice Boltzmann (LB) method  derived from kinetic theory. The advantage of LB method lies in its inherent ability to capture interfacial evolution and charge transport at the mesoscopic level without explicit interface tracking. Moreover, LB method excels in handling complex geometric boundaries and is highly amenable to large-scale parallel computing \cite{Lee:JCP2010,Liang:PRE2016,Wang:C2019,Liu:MMS2022}, making it particularly suitable for 
complex flows involving multiphase interfaces and multiphysics coupling. Because of these strengths, the phase-field LB method has been extensively adopted for EHD multiphase flow research. Recent studies have explored complex physical interfacial phenomena, such as the collision dynamics of charged droplets \cite{Shardt:CEJ2016} and nucleate boiling cycles \cite{Al-Sadik:ICHMT2025}. To further enhance the method, researchers have developed regularized schemes for accuracy and stability \cite{Zhu:POF2025} and proposed mass-conserving frameworks for complex multiphase systems \cite{Liu:PD2024}.  Despite these advancements, these works remain grounded in phenomenological models and generally neglect the critical issues of thermodynamic consistency and surface charge convection. In this work, we address these gaps by pursuing three primary objectives: (i) to derive a thermodynamically consistent three-phase EHD model built upon the Onsager variational principle, effectively capturing surface charge convection; (ii) to establish an efficient mesoscopic LB method framework for its numerical solution; and (iii) to validate the framework using benchmarks and apply it to simulate complex ternary EHD phenomena.

The paper is structured as follows.  \Cref{sec:variation} details the rigorous theoretical derivation of the proposed physical model.  \Cref{sec:LBM} outlines the corresponding LB numerical implementation. In  \cref{sec:experiments}, the precision and dependability of the proposed computational framework are initially demonstrated via a succession of benchmark tests, including electroosmotic flow in microchannels, the spreading of a three-phase liquid lens, and the static equilibrium as well as the dynamic deformation of compound droplets; subsequently, it applies the validated framework to investigate complex EHD phenomena, specifically focusing on the coalescence and separation dynamics of double droplets under an electric field, and the droplet behavior subjected to combined EHD and shear flows. Finally, \cref{sec:conclusion} summarizes the main conclusions.
\section{A variational framework for three-phase EHD}
\label{sec:variation}
\subsection{Onsager’s Variational Principle}
The Onsager variational principle \cite{Onsager:IPR1931} is a thermodynamic foundation for modeling irreversible transport and dissipative systems. Based on microscopic reversibility, it yields thermodynamically consistent governing equations by balancing driving forces and energy dissipation.

For isothermal multiphysics systems, this classic entropy foundation naturally transitions into  free energy framework. The system's evolution is governed by the total free energy, denoted as $\mathcal{F}(\boldsymbol{\alpha})$. The rate of change of the free energy is expressed through the chain rule as:
\begin{equation}
	\dot{\mathcal{F}}(\boldsymbol{\alpha}; \dot{\boldsymbol{\alpha}}) =  \frac{\partial \mathcal{F}}{\partial \alpha_i} \dot{\alpha}_i.
	\label{eq1} 
\end{equation}

Based on the structural symmetry established by Onsager's classical theory ($\zeta_{ij}=\zeta_{ji}$), the energy dissipation is  quantified by a dissipation function $\Phi_{\mathcal{F}}$,which is a positive-definite quadratic functional of the evolution rates:
\begin{equation}
	\Phi_{\mathcal{F}}(\boldsymbol{\alpha}; \dot{\boldsymbol{\alpha}}) = \frac{1}{2}  \zeta_{ij} \dot{\alpha}_i \dot{\alpha}_j,
	\label{eq2} 
\end{equation}
where $\zeta_{ij}$ is the friction coefficient. The theoretical foundation is established by a unified variational objective function, denoted as the Rayleighian $\mathcal{R}$ \cite{Xu:JPCM2015,Chen:JCP2025}:
\begin{equation}
	\mathcal{R}(\boldsymbol{\alpha}; \dot{\boldsymbol{\alpha}}) = \dot{\mathcal{F}} + \Phi_{\mathcal{F}} =  \frac{\delta \mathcal{F}}{\delta \alpha_i} \dot{\alpha}_i + \frac{1}{2}  \zeta_{ij} \dot{\alpha}_i \dot{\alpha}_j.
	\label{eq3} 
\end{equation}

The variational principle postulates that the actual physical trajectory of the system's evolution is determined by minimizing the Rayleighian with respect to the rates $\dot{\alpha}_i$. The minimization condition, $\delta \mathcal{R} / \delta \dot{\alpha}_i = 0$, yields the force balance equation:\begin{equation}
	\zeta_{ij} \dot{\alpha}_j + \frac{\delta \mathcal{F}}{\delta \alpha_i} = 0,
	\label{eq4} 
\end{equation}
which establishes a fundamental relationship governing the  evolution of the state variables.

This  formulation ensures that the resulting dynamics strictly adhere to the second law of thermodynamics, guaranteeing a monotonically decreasing free energy over time ($\dot{\mathcal{F}} \le 0$).  The variational principle is a powerful tool for deriving constitutive equations for coupled systems \cite{Eck:IFB2009,Xiong:jsc2025,Doi:JPCM2011}, well-suited for three-phase EHD problems.

\subsection{The Generalized Governing Equations }
Based on Onsager’s variational principle, this section formulates the governing equations for a three-phase EHD system. By minimizing the Rayleighian, the unknown fluxes and forces are determined, ensuring that the resulting  relations satisfy the second law of thermodynamics and  embed multiphysics couplings within a unified variational framework.

We consider an isothermal, incompressible system of three immiscible Newtonian phases coupled via hydrodynamic and electrodynamic interactions. Domain boundaries are impermeable and adiabatic. Three order parameters $\phi_1$, $\phi_2$, and $\phi_3$ denote the volume fractions of each phase, subject to the constraint \cite{Boyer:ESAIM2006,Boyer:TPM2010}:
\begin{equation}
\phi_1+\phi_2+\phi_3=1,\quad0\leq\phi_i\leq1.
	\label{eq5} 
\end{equation}
The overall state of  system is determined by the total free energy functional ($\mathcal{F}$).  It  comprises four contributions:  chemical free energy ($\mathcal{F}_\phi$),  electrostatic energy ($\mathcal{F}_\mathbf{E}$),  macroscopic kinetic energy ($\mathcal{F}_\mathbf{u}$), and  charge diffusion  energy ($\mathcal{F}_q$) \cite{Eck:IFB2009}:
\begin{equation}
	\mathcal{F}=\mathcal{F}_\phi+\mathcal{F}_\mathbf{E}+\mathcal{F}_\mathbf{u}+\mathcal{F}_q
	\label{eq6} 
\end{equation}

Based on Landau theory, the free energy functional over domain 
$\Omega$  is constructed as the sum of a bulk free energy density and gradient terms  \cite{Boyer:ESAIM2006,Boyer:TPM2010}:
\begin{equation}
	\mathcal{F}_\phi=\int_\Omega\left[\frac{12}{D}F(\phi_1,\phi_2,\phi_3)+\sum_{i=1}^3\frac{3}{8}D\lambda_i|\nabla\phi_i|^2\right]d\Omega,
	\label{eq7} 
\end{equation}
where $D$ denotes the characteristic interface thickness. $F(\phi_1,\phi_2,\phi_3)$ is the bulk free energy density, expressed as
\begin{equation}
	F(\phi_1,\phi_2,\phi_3)=\left[\frac{\lambda_1}{2}\phi_1^2(1-\phi_1)^2+\frac{\lambda_2}{2}\phi_2^2(1-\phi_2)^2+\frac{\lambda_3}{2}\phi_3^2(1-\phi_3)^2\right].
	\label{eq8} 
\end{equation}
Here,  the  positive parameters $\lambda_{i}$ related to the interfacial tensions, are defined as
\begin{equation}
	\begin{aligned}
		\lambda_1=\gamma_{12}+\gamma_{13}-\gamma_{23},\quad
		\lambda_2=\gamma_{12}+\gamma_{23}-\gamma_{13},\quad
		\lambda_3=\gamma_{13}+\gamma_{23}-\gamma_{12},
	\end{aligned}
	\label{eq9} 
\end{equation}
where  $\gamma_{12}$, $\gamma_{13}$ and $\gamma_{23}$ are the surface tension between each pair of fluid phases in the ternary system. Under an applied electric field, the electrostatic free energy is \cite{Zhang:JCP2024}:
\begin{equation}
	\mathcal{F}_{\mathbf{E}}=\frac{1}{2}\int_\Omega\frac{|\mathbf{D}|^2}{\varepsilon(\phi_1,\phi_2,\phi_3)}d\Omega,
	\label{eq10} 
\end{equation}
where $\mathbf{D}=\varepsilon(\phi_1,\phi_2,\phi_3)\mathbf{E}$ is the electric displacement field, and the permittivity $\varepsilon$ is a function of the phase-field variables. The electric field strength $\boldsymbol{E}$ is defined as the negative gradient of the electrostatic potential, yielding $\boldsymbol{E} = -\nabla\psi$.  Moreover, the macroscopic kinetic energy resulting from the fluid motion is expressed as \cite{Zhang:JCP2024}:
\begin{equation}
	\mathcal{F}_{\mathbf{u}}=\frac{1}{2}\int_\Omega\rho|\mathbf{u}|^2d\Omega,
	\label{eq11} 
\end{equation}
where $\rho$ denotes the density and $\mathbf{u}$ is the fluid velocity. Additionally, the free energy contribution associated with charge diffusion is incorporated as \cite{Eck:IFB2009}:
\begin{equation}
	\mathcal{F}_{q}=\frac{\lambda}{2}\int_\Omega q^2d\Omega,
	\label{eq12} 
\end{equation}
where $q$ indicates the free charge density, and $\lambda$ denotes a coupling coefficient reflecting the charge transport characteristics.

Based on the three-phase EHD system's total free energy functional, the temporal evolution of the phase-field variables follows the convective Cahn–Hilliard equations:\begin{equation}
	\frac{\partial\phi_i}{\partial t}+\mathbf{u}\cdot\nabla\phi_i=-\nabla\cdot\mathbf{J}_{\phi_i},\quad i=1,2,3
	\label{eq13} 
\end{equation}
where $\mathbf{J}_{\phi_i}$ denotes the  flux of the $i$-th phase. The electrostatic field satisfies the Maxwell equations \cite{Landau:PO1975}
\begin{equation}
	\nabla\cdot(\varepsilon\mathbf{E})=q,\quad\mathbf{E}=-\nabla\psi.
	\label{eq14} 
\end{equation}
The  hydrodynamics are given by the incompressible Navier–Stokes (NS) equations \cite{Castellanos:S2014},
\begin{equation}
	\begin{aligned}
		&\nabla\cdot\mathbf{u}=0,\\&\rho\left(\frac{\partial\mathbf{u}}{\partial t}+\mathbf{u}\cdot\nabla\mathbf{u}\right)=\nabla\cdot\mathbf{\Pi}+\mathbf{F},
	\end{aligned}
	\label{eq15} 
\end{equation}
with the  stress tensor $\mathbf{\Pi}$ formulated as:
\begin{equation}
	\mathbf{\Pi}=-p\mathbf{I}+\mu\left[\nabla\mathbf{u}+\left(\nabla\mathbf{u}\right)^\mathrm{T}\right],
	\label{eq16} 
\end{equation}

The body force $\mathbf{F}$ combines interfacial and electrostatic forces, while free charge conservation follows the convection–conduction equation:
\begin{equation}
	\frac{\partial q}{\partial t}+\mathbf{u}\cdot\nabla q+\nabla\cdot\mathbf{J}_D=0,
	\label{eq17} 
\end{equation}
with $\mathbf{J}_D$ denoting the conduction current density.

In the above governing equations, the phase fluxes $\mathbf{J}_{\phi_i}$ , conduction current 
$\mathbf{J}_D$ , and body force $\mathbf{F}$ are unknown, leaving the system unclosed. To close it, we employ Onsager’s variational principle. Selecting $\mathbf{u}$, $\mathbf{J}_{\phi_i}$, and $\mathbf{J}_D$  as independent fluxes, we construct the Rayleighian from the total free energy variation rate and a dissipation potential . Minimizing the Rayleighian yields constitutive relations for $\mathbf{J}_{\phi_i}$, $\mathbf{J}_D$ , and  $\mathbf{F}$ that satisfy the second law. Together with the conservation equations, these form a complete, thermodynamically consistent model for three-phase EHD flows.

From Eqs. (\ref{eq6}) –(\ref{eq12}), the total energy of the system can be formulated as:
\begin{equation}
	\begin{aligned}
		\mathcal{F}&=\int_\Omega\sum_{i=1}^3\left[\frac{6}{D}\lambda_i\phi_i^2(1-\phi_i)^2+\frac{3}{8}D\lambda_i|\nabla\phi_i|^2\right]d\Omega\\
		&+\frac{1}{2}\int_\Omega\frac{|\mathbf{D}|^2}{\varepsilon(\phi_1,\phi_2,\phi_3)}d\Omega	+\frac{1}{2}\int_\Omega\rho|\mathbf{u}|^2d\Omega+\frac{\lambda}{2}\int_\Omega q^2d\Omega.
	\end{aligned}
	\label{eq18} 
\end{equation}
By taking the variation of the total free energy, it follows that:
\begin{equation}
	\delta\mathcal{F}=\int_{\Omega}\sum_{i=1}^3[\frac{24}{D}\lambda_i\phi_i(\phi_i-\frac{1}{2})(\phi_i-1)-\frac{3}{4}D\lambda_i\nabla^{2}\phi)\delta\phi_i-\frac{\varepsilon^{\prime}(\phi_i)}{2\varepsilon^2(\phi_i)}|\mathbf{D}|^2 ]d\Omega
	\label{eq19} 
\end{equation}
where the Gauss divergence theorem has been utilized. Minimizing $\mathcal{F}$ with respect to the order parameter $\phi_i$ yields the following equilibrium conditions:
\begin{equation}
	\frac{24}{D}\lambda_i\phi_i(\phi_i-\frac{1}{2})(\phi_i-1)-\frac{3}{4}D\lambda_i\nabla^{2}\phi
	-\frac{\varepsilon^{\prime}(\phi)}{2\varepsilon^2(\phi)}|\boldsymbol{D}|^2=\mu_i\equiv\mathrm{const},\quad\mathrm{~in~}\Omega,
	\label{eq20} 
\end{equation}
where $\mu_i$ denotes the chemical potential of the $i$-th phase. In light of Eq. (\ref{eq18}), the temporal variation of the total free energy can be formulated as:
\begin{equation}
	\dot{\mathcal{F}}=\int_\Omega{\sum_{i=1}^3[\mu_i\frac{\partial\phi_i}{\partial t}]}d\Omega+\int_\Omega{(\mathbf{E}\cdot\frac{\partial\mathbf{D}}{\partial t})}d\Omega+\int_\Omega{(\rho\boldsymbol{u}\cdot\frac{\partial\boldsymbol{u}}{\partial t})}d\Omega+\lambda\int_\Omega{(q\frac{\partial q}{\partial t})}d\Omega.
	\label{eq21} 
\end{equation}
Combining Eqs. (\ref{eq13}) and (\ref{eq17}), and employing integration by parts, we find that:
\begin{equation}
	\begin{aligned}
		\int_{\Omega}\left[\mu_i\left(\frac{\partial\phi_i}{\partial t}+\nabla\cdot\phi_i\mathbf{u}\right)\right]d\Omega&=-\int_{\Omega}\mu_i\nabla\cdot\mathbf{J}_{\phi_i}d\Omega=\int_{\Omega}\nabla\mu_i\cdot\mathbf{J}_{\phi_i}d\Omega,\\
	\int_{\Omega}\left[q\left(\frac{\partial q}{\partial t}+\mathbf{u}\cdot\nabla q\right)\right]d\Omega	&=-\int_{\Omega}q\nabla\cdot\mathbf{J}_{D}d\Omega=\int_{\Omega}\nabla q\cdot\mathbf{J}_{D}d\Omega,
	\end{aligned}
	\label{eq22} 
\end{equation}
where  impermeable boundaries $\mathbf{n}_w\cdot\mathbf{J}_{\phi_i}=0$, and $\mathbf{n}_w\cdot\mathbf{J}_D=0$ have been enforced at the solid surfaces.  Submitting Eq.  (\ref{eq22}) into Eq.  (\ref{eq21}), the energy rate becomes:
\begin{equation}
	\begin{aligned}
		&\dot{\mathcal{F}}=\sum_{i=1}^3\int_{\Omega}[\nabla\mu_i\cdot\mathbf{J}_{\phi_i}+\phi_i\mathbf{u}\cdot\nabla\mu_i]d\Omega+\int_{\Omega}\mathbf{E}\cdot(-q\mathbf{u}-\mathbf{J}_{\mathbf{D}})d\Omega\\+&\int_{\Omega}(\boldsymbol{u}\cdot\mathbf{F}-\Pi:\nabla\mathbf{u})d\Omega+\int_{\Omega}[\mathbf{n}_{\mathbf{w}}\cdot(\Pi\mathbf{u}_{\tau})]dA+\lambda\int_{\Omega}[\nabla q\cdot\mathbf{J}_{\mathbf{D}}-q\boldsymbol{u}\cdot\nabla q]d\Omega,
	\end{aligned}
	\label{eq23}
\end{equation}
where $\mathbf{\tau}$ indicates the tangential direction along the solid boundary. 

Assuming  no-slip boundary at the solid walls for a system deviating from thermodynamic equilibrium, the total energy dissipation within the system originates from viscous effects The dissipation function $\Phi_F(\dot{\alpha}_i,\dot{\alpha})$ takes the following form:
\begin{equation}
	\Phi_F=\int_\Omega\frac{|\boldsymbol{\Pi}|^2}{2\mu}d\Omega+\int_{\Omega}\frac{1}{2}\sum_{i=1}^{3}\sum_{j=1}^{3}\mathbf{J}_{\phi_{i}}\cdot M^{-1}_{ij}\mathbf{J}_{\phi_{j}}d\Omega+\int_\Omega\frac{|\mathbf{J}_D|^2}{2\sigma}d\Omega
	\label{eq24}
\end{equation}
where $M_{ij}$ re  matrix elements governing diffusion between phases  $i$ and  $j$. This formulation  links energy dissipation to shear viscosity (first term), multi-component diffusion (second term), and electrical conduction (third term) in the bulk. By combining   Eq. (\ref{eq23}) and Eq.  (\ref{eq24}), we construct the Rayleighian functional $\mathcal{R} = \dot{\mathcal{F}} + \Phi_F$:
\begin{equation}
	\begin{aligned}
		\mathcal{R}&=\sum_{i=1}^3\int_{\Omega}[\nabla\mu_i\cdot\mathbf{J}_{\phi_i}+\phi_i\mathbf{u}\cdot\nabla\mu_i]d\Omega+\int_{\Omega}\mathbf{E}\cdot(-q\mathbf{u}-\mathbf{J}_{\mathbf{D}})d\Omega\\&+\int_{\Omega}(\boldsymbol{u}\cdot\mathbf{F}-\Pi:\nabla\mathbf{u})d\Omega+\int_{\Omega}[\mathbf{n}_{\mathbf{w}}\cdot(\Pi\mathbf{u}_{\tau})]dA+\lambda\int_{\Omega}[\nabla q\cdot\mathbf{J}_{\mathbf{D}}-q\boldsymbol{u}\cdot\nabla q]d\Omega\\&+\int_\Omega\frac{|\boldsymbol{\Pi}|^2}{2\mu}d\Omega+\int_{\Omega}\frac{1}{2}\sum_{i=1}^{3}\sum_{j=1}^{3}\mathbf{J}_{\phi_{i}}\cdot M^{-1}_{ij}\mathbf{J}_{\phi_{j}}d\Omega+\int_\Omega\frac{|\mathbf{J}_D|^2}{2\sigma}d\Omega.
	\end{aligned}
	\label{eq25}
\end{equation}

Subsequently, minimizing $\mathcal{R}$ with respect to the independent generalized fluxes ($\mathbf{u}$, $\mathbf{J}_{\phi_i}$, and $\mathbf{J}_D$), subject to the incompressibility constraint ($\nabla\cdot\mathbf{u}=0$), yields the requisite constitutive relations:
\begin{equation}
	\begin{aligned}
		\mathbf{F}=-\sum_{i=1}^3\phi_i\nabla\mu_i+q\mathbf{E}-\frac{1}{2}\nabla\left(\frac{\alpha}{\sigma}q^2\right),\quad
		\mathbf{J}_{\phi_i}=-\sum_{i=1}^{3}M_{ij}\nabla\mu_{j},\quad\mathbf{J}_{D}=\sigma(\mathbf{E}-\lambda\nabla q).
	\end{aligned}
	\label{eq26}
\end{equation}

By substituting the constitutive relations from Eq. (\ref{eq26})  back into the generalized macroscopic equations, we establish the  coupled, thermodynamically consistent governing equations for three-phase EHD flows:

\begin{subequations}
	\begin{align}
	&	\nabla\cdot\mathbf{u} = 0, \label{eq27a} \\
	&	\rho\left(\frac{\partial\mathbf{u}}{\partial t}+\mathbf{u}\cdot\nabla\mathbf{u}\right) 
		= -\nabla p + \nabla\cdot[\mu(\nabla\mathbf{u}+\nabla\mathbf{u}^{\mathrm{T}})] 
		- \sum_{i=1}^3 \phi_i\nabla\mu_i + q\mathbf{E}, \label{eq27b} \\
		\frac{\partial\phi_i}{\partial t} + \nabla\cdot(\phi_i\mathbf{u}) &
		= \sum_{j=1}^{3} \nabla\cdot(M_{ij}\nabla\mu_j),\quad 
		\mu_i = \frac{24}{D}\lambda_i\phi_i^2(1-\phi_i)^2 + \frac{3}{4}D\lambda_i|\nabla\phi_i|^2 
		- \frac{1}{2}\varepsilon'(\phi_i)|\mathbf{E}|^2, \label{eq27c} \\
	&	\nabla\cdot(\varepsilon\mathbf{E}) = q,\quad \mathbf{E} = -\nabla\varphi, \label{eq27d} \\
	&	\frac{\partial q}{\partial t} + \mathbf{u}\cdot\nabla q = \nabla\cdot\bigl[\alpha\nabla q + \sigma\nabla\varphi\bigr]. \label{eq27e}
	\end{align}
\label{eq27}
\end{subequations}

To guarantee reduction consistency, the three-phase model is required to recover the standard two-phase system when the  $i$-th fluid is absent. This requires the mobility tensor $M_{ij}=0$ when $\phi_i = 0$, leading to the following definition:
\begin{equation}
	M_{ij}=
	\begin{cases}
		-m_0\phi_i\phi_j,&i\neq j,\\
		-\sum_{j,j\neq i}M_{ij},&i=j.
	\end{cases}
	\label{eq28}
\end{equation}

\textit{Remark}: Based on the generalized Rayleighian formulation in Eq. (\ref{eq25}) and the resulting governing Eqs. (\ref{eq27}), it is noteworthy that the chemical potential $\mu_i$ contains the explicit electrical coupling term $-\frac{1}{2}\varepsilon'(\phi_i)|\mathbf{E}|^2$. This suggests that the conventional multi-component chemical potential, derived solely from the Ginzburg-Landau bulk energy, is theoretically insufficient. Unlike traditional phenomenological models, the present study ensures that the electrical energy density is an intrinsic part of the multi-phase thermodynamic equilibrium. Furthermore, by adopting the mobility matrix $M_{ij}$ defined in Eq. (\ref{eq28}), our model ensures reduction-consistency that  when any phase $\phi_i$ vanishes, the system rigorously recovers the two-phase EHD equations supported by Eck et al. \cite{Eck:IFB2009} and Xiong et al. \cite{Xiong:jsc2025}. Therefore, this work establishes a thermodynamically  phase‑field formulation for three‑phase EHD flows.

\subsection{Physical Property Interpolation}
In the diffuse-interface framework, macroscopic properties $\rho$, $\mu$, $\sigma$ are continuous functions of $\phi_1, \phi_2$, and $\phi_3$. A linear volume-averaging rule is applied to  these properties, ensuring exact recovery of bulk properties in each pristine phase:
\begin{equation}
	\rho= \sum_{i=1}^{3} \phi_i \rho_i,\quad
	\mu = \sum_{i=1}^{3} \phi_i \mu_i, \quad
	\sigma = \sum_{i=1}^{3} \phi_i \sigma_i,
	\label{eq29}
\end{equation}
where $\rho_i$, $\mu_i$, and $\sigma_i$ represent the constant physical parameters of the $i$-th fluid phase. For electric permittivity $\varepsilon$, higher-order interpolation is required to ensure numerical stability, requiring $\partial\varepsilon/\partial\phi_i = 0$  at pure phases. A smoothed Hermite interpolation $H(\phi_i)$ is thus introduced: $H(\phi_i) = \phi_i^2 (3 - 2\phi_i).$

The global permittivity is then assembled by superimposing the individual phase contributions, weighted by this nonlinear smoothing function:
\begin{equation}
	\varepsilon= \sum_{i=1}^{3} \varepsilon_i H(\phi_i) = \sum_{i=1}^{3} \varepsilon_i \phi_i^2 (3 - 2\phi_i).
	\label{eq30}
\end{equation}
which ensures the electric body force remains well-behaved across the interface regions.

\section{Lattice Boltzmann method}
\label{sec:LBM}
Diverging from traditional macroscopic computational fluid dynamics techniques, the LB method is rooted in kinetic theory. It characterizes fluid dynamics by tracking the evolution of particle distribution functions across a discretized velocity space. Velocity sets are denoted as $DdQq$ lattices ($d$: dimensions, $q$: velocities) \cite{Kruger:SIP2017,Liang:PRE2014}, e.g., $D2Q9$ and $D3Q15$:

$D2Q9$:
\begin{equation}
	\begin{aligned}	
		\mathbf{c}_i=&\begin{bmatrix}0&1&0&-1&0&1&-1&-1&1\\0&0&1&0&-1&1&1&-1&-1\end{bmatrix}c,\\
		&\omega_{0}=4/9,\omega_{1-4}=1/9,\omega_{5-8}=1/36;
	\end{aligned}
	\label{eq31}
\end{equation}

$D3Q15$:
\begin{equation}
	\begin{aligned}	
		\mathbf{c}_i=&\left[\begin{array}{rrrrrrrrrrrrrrr}0&1&-1&0&0&0&0&1&-1&1&-1&1&-1&-1&1\\0&0&0&1&-1&0&0&1&-1&1&-1&-1&1&1&-1\\0&0&0&0&0&1&-1&1&-1&-1&1&1&-1&1&-1\end{array}\right]c,\\
		&\omega_{0}=2/9,\omega_{1-6}=1/9,\omega_{7-14}=1/72,
	\end{aligned}
	\label{eq32}
\end{equation}
In these expressions, $\omega_i$ denotes the weighting factors and $\mathbf{c}_i$ corresponds to the discrete velocity vectors. The lattice velocity is given by $c=\delta_{x}/\delta_{t}$, where $\delta x$ represents the mesh spacing and $\delta t$ is  the time step, yielding a lattice speed of sound $c_s=c /\sqrt{3}$.

The standard LB algorithm consists of collision and streaming steps. We adopt the single-relaxation-time (SRT) collision operator \cite{He:PRE1997} for its computational efficiency and simplicity. Spatial derivatives in the LB formulation are discretized using second-order isotropic finite difference schemes \cite{Lou:EL2012}:
\begin{equation}
		\nabla\chi\left(\mathbf{x},t\right)=\sum_{i}\frac{\omega_{i}\mathbf{c}_{i}\chi\left(\mathbf{x}+\mathbf{c}_{i}\delta_{t},t\right)}{c_{s}^{2}\delta_{t}}, 
		\nabla^{2}\chi\left(\mathbf{x},t\right)=\sum_{i}\frac{2\omega_{i}[\chi\left(\mathbf{x}+\mathbf{c}_{i}\delta_{t},t\right)-\chi\left(\mathbf{x},t\right)]}{c_{s}^{2}\delta_{t}^{2}},	
	\label{eq33}
\end{equation}
where $\chi$ represents an arbitrary variable.

\subsection{ Lattice Boltzmann Method for Cahn-Hilliard Equation}

Directly applying the standard LB framework to the CH equation often triggers numerical instability because the mobility tensor $M_{ij}$ is  dependent on the order parameter $\phi_i$. To circumvent this issue, we partition the diffusion flux $M_{ii}\nabla\mu_{i}$ into two distinct components: $m_0\nabla\mu_{i}$ and $(M_{ii}-m_0)\nabla\mu_{i}$. The former, featuring a constant  mobility $m_0$, acts as a pure diffusion term, whereas the remainder is incorporated as an external source term \cite{Liu:SIAMJSC2025}. The discrete evolution for the $i$-th phase ($i=1, 2, 3$) distribution function, $f_j^i$, augmented with this source term, reads :\begin{equation}
	f_{j}^{i}(\mathbf{x}+\mathbf{c}_{j}\Delta t,t+\Delta t)=f_{j}^{i}(\mathbf{x},t)-\frac{1}{\tau_{f}}[f_{j}^{i}(\mathbf{x},t)-f_{j}^{i,eq}(\mathbf{x},t)]+\Delta t(1-\frac{1}{2\tau_{f}})F_{j}^{i}(\mathbf{x},t),
	\label{eq34}
\end{equation}
Here, the relaxation time $\tau_{f}$ is  coupled with the reference mobility $m_0$ according to: $m_0=c_s^2(\tau_{f}-0.5)\Delta t,$ The equilibrium distribution function $f_j^{i,eq}$  is formulated as:
\begin{equation}
	f_j^{i,eq}\left(\mathbf{x},t\right)=
	\begin{cases}
		\phi_i-\left(1-\omega_0\right)\mu_i,&j=0,\\
		\omega_j\mu_i+\omega_j\frac{\mathbf{c}_j\cdot\phi_i \bf{u}}{c_s^2},&j\neq0.
	\end{cases}
	\label{eq35}
\end{equation}
with $\mu_i$ denoting the local chemical potential for phase $i$. The discrete source term $F_j^i(\mathbf{x},t)$ is explicitly formulated as:
\begin{equation}
	F_{j}^{i}(\mathbf{x},t)=\omega_j\left[\frac{\mathbf{c}_j\cdot\partial_t( \phi_i\mathbf{u})}{c_s^2}-\left(\frac{ M_{ii}- m_{0}}{m_{0}}\nabla\mu_{i}+\sum_{k\neq i}\frac{ M_{ik}}{m_{0}}\nabla\mu_{k}\right)\right],.
	\label{eq36}
\end{equation}
The macroscopic order parameter for each phase is calculated by the zeroth moment: $\phi_i=\sum_jf_{j}^{i,eq}.$

\subsection{ Lattice Boltzmann Method for  Hydrodynamic Equations}

The hydrodynamic field is resolved via the distribution function  $g_j$
. The evolution equation incorporating the total force term is defined as \cite{Liang:PRE2014,Liang:PRE2018}:
\begin{equation}
	g_j(\mathbf{x}+\mathbf{c}_j\Delta t,t+\Delta t)=g_j(\mathbf{x},t)-\frac{1}{\tau_g}[g_j(\mathbf{x},t)-g_j^{eq}(\mathbf{x},t)]+\Delta t(1-\frac{1}{2\tau_g})\bar{G}_j(\mathbf{x},t),
	\label{eq37}
\end{equation}

The formulation for the forcing term $\bar{G}_j$ is given by:
\begin{equation}
	\bar{G}_j(\mathbf{x},t)=\omega_j[\mathbf{u}\cdot\nabla\rho+\frac{\mathbf{c}_j\cdot\mathbf{F}}{c_s^2}+\frac{\mathbf{u}\nabla\rho:(\mathbf{c}_j\mathbf{c}_j-c_s^2\mathbf{I})}{c_s^2}].
	\label{eq38}
\end{equation}
Meanwhile, the equilibrium distribution $g_j^{eq}$ is constructed as:
\begin{equation}
	g_j^{eq}(\mathbf{x},t)=
	\begin{cases}
		&(\omega_0-1)\frac{p}{c_s^2}+\rho s_j(\mathbf{u}),j=0,\\
		&\omega_j\frac{p}{c_s^2}+\rho s_j(\mathbf{u}),\quad j\neq0,
	\end{cases}
	\label{eq39}
\end{equation}
where the velocity-dependent function $s_j(\mathbf{u})$ is defined as:\begin{equation}
	s_j(\mathbf{u})=\omega_j[\frac{\mathbf{c}_j\cdot\mathbf{u}}{c_s^2}+\frac{(\mathbf{c}_j\cdot\mathbf{u})^2}{2c_s^4}-\frac{\mathbf{u}\cdot\mathbf{u}}{2c_s^2}].
	\label{eq40}
\end{equation}
The fluid kinematic viscosity $\mu$ determines the relaxation time via $\mu=\rho c_s^2\left(\tau_g-0.5\right).$ 

The macroscopic velocity  $\mathbf{u}$ and pressure $p$ are obtained as:
\begin{equation}
	\begin{aligned}
		\rho\mathbf{u}=\sum_j\mathbf{c}_jg_j+\frac{1}{2}\Delta t\mathbf{F},\quad
		p=\frac{c_s^2}{1-\omega_0}[\sum_{j\neq0}g_i+\frac{\Delta t}{2}\mathbf{u}\cdot\nabla\rho+\rho s_0(\mathbf{u})].
	\end{aligned}
	\label{eq41}
\end{equation}

\subsection{ Lattice Boltzmann Method for Electric Potential Equation}

The spatial configuration of the electric potential is determined by the Poisson equation, accommodating  varying permittivity:
\begin{equation}
	\nabla\cdot\varepsilon\left(\phi\right)\nabla\varphi+q=0.
	\label{eq46}
\end{equation}
In conventional LB frameworks, the phase-dependent permittivity $\varepsilon(\phi)$ is directly tied to the  relaxation time $\tau_\varphi$ via, $\varepsilon\left(\phi\right)= c_s^2\left(\tau_\varphi-0.5\right)\Delta t.$

Although viable for uniform or low-contrast diffusion problems,this scheme suffers from severe computational inefficiency and numerical instability in EHD flows. This limitation arises because large permittivity ratios induce extreme disparities in $\tau_\varphi$ across the interface \cite{Xiong:jsc2025}. To circumvent these computational bottlenecks, we decompose the variable permittivity, reformulating the equation as:
\begin{equation}
	\nabla\cdot\varepsilon_3\nabla\varphi+\nabla\cdot\mathbf{\hat{\varepsilon}}\nabla\varphi+q=0,
	\label{eq42}
\end{equation}
where $\hat{\varepsilon}=\varepsilon-\varepsilon_3$
. The dynamic progression of the potential distribution function $h_j$ is governed by:
\begin{equation}
	h_j(\mathbf{x}+\mathbf{c}_j\Delta t,t+\Delta t)=h_j(\mathbf{x},t)-\frac{1}{\tau_h}[h_j(\mathbf{x},t)-h_j^{eq}(\mathbf{x},t)]-\frac{\hat{\varepsilon}}{c_s^2}\frac{\omega_j\mathbf{c}_j\cdot\nabla\varphi}{\tau_h}+\Delta t\varpi_jq,
	\label{eq43}
\end{equation}
where the weight coefficients are given by $\varpi_0=0,\varpi_{1-8}=1/8$. The relaxation time is set as  $\tau_h = 0.5 + \varepsilon_3 / (c_s^2 \Delta t)$, while the equilibrium function $h_j^{eq}$ is specified by:
\begin{equation}
	h_j^{eq}(\mathbf{x},t)=
	\begin{cases}
		(\omega_0-1)\varphi(\mathbf{x},t),j=0,\\
		\omega_j\varphi(\mathbf{x},t),\quad j\neq0,
	\end{cases}
	\label{eq44}
\end{equation}

The electric potential $\varphi$ and its gradient are calculated locally:
\begin{equation}
	\begin{aligned}
		\varphi=\sum_{j\neq0}\frac{1}{1-\omega_0}h_j,
		\nabla\varphi=-\frac{\sum_j\mathbf{c}_jh_j}{\hat{\varepsilon}+c_s^2\tau_h\Delta t}.
	\end{aligned}
	\label{eq45}
\end{equation}

\subsection{ Lattice Boltzmann Method for Nernst-Planck Equation}
The numerical resolution of the Nernst-Planck equation presents inherent challenges, primarily due to the system's nonlinearity arising from the strong coupling between the electric field and the charge density. Moreover, charges are highly confined within a thin transition region around the interface, necessitating the precise capture of sharp charge density gradients across a remarkably narrow zone. To facilitate the numerical simulation, a diffuse-interface strategy is employed to resolve the localized charge density. Concurrently, a mathematical transformation is applied to incorporate the Ohmic conduction effect directly as a  source term. Consequently, the governing equation for charge transport is reformulated into the following equivalent expression \cite{Luo:POF2020}:\begin{equation}
	\frac{\partial q}{\partial t}+\mathbf{u}\cdot\nabla q=\nabla\cdot(\alpha\nabla q)+R.
	\label{eq46}
\end{equation}
The forcing term $R$ is defined as
\begin{equation}
	R=-\frac{\sigma q}{\varepsilon}+\frac{\sigma}{\varepsilon}\nabla\varepsilon\cdot\mathbf{E}-\nabla\sigma\cdot\mathbf{E},
	\label{eq47}
\end{equation}
To simulate this transport phenomenon, the distribution function $l_j$ is given by \cite{Chai:JSC2016}:
\begin{equation}
	l_j(\mathbf{x}+\mathbf{c}_j\Delta t,t+\Delta t)=l_j(\mathbf{x},t)-\frac{1}{\tau_l}[l_j(\mathbf{x},t)-l_j^{eq}(\mathbf{x},t)]+\Delta tS_j(\mathbf{x},t)+\Delta tT_j(\mathbf{x},t),
	\label{eq48}
\end{equation}
The source terms $S_j$ and $T_j$ are mathematically evaluated as:
\begin{equation}
	\begin{aligned}
		S_{j}(\mathbf{x},t)=(1-\frac{1}{2\tau_l})\omega_jR,\quad
		T_{j}(\mathbf{x},t)=(1-\frac{1}{2\tau_l})\frac{\omega_j\mathbf{c}_j\cdot\partial_t(q\mathbf{u})}{c_s^2}.
	\end{aligned}
	\label{eq49}
\end{equation}
with the relaxation time given by $\tau_l=0.5+\alpha/c_s^2\Delta t.$. The equilibrium distribution function $l_j^{eq}$
is given by:
\begin{equation}
	l_j^{eq}(\mathbf{x},t)=\omega_jq(1+\frac{\mathbf{c}_j\cdot\mathbf{u}}{c_s^2}).
	\label{eq50}
\end{equation}
Finally, the charge density  $q$ is computed by: $q=\sum_jl_j+0.5\Delta tR$.

\section{Numerical solutions}
\label{sec:experiments}
This section validates the proposed LB framework through several benchmarks: electroosmotic flow in microchannels, static three-phase droplets, spreading of a three-phase liquid lens, and compound droplet deformation under a uniform electric field. The validated model then investigates complex EHD behavior of multiphase systems, including double droplet coalescence and separation under an electric field, as well as droplet morphology under combined EHD and shear flow. Key factors such as permittivity and conductivity ratios are also considered.

The computational results are presented using non-dimensional parameters. Here we simply outline the relevant dimensionless numbers, defined as follows \cite{Santra:POF2020}:
\begin{equation}
	\begin{aligned}
		&Ca=\frac{\mu_{3}\dot{\gamma}_{0}R_{b}}{\gamma_{23}},\quad Re=\frac{\rho_{3}\dot{\gamma}_{0}R_{b}^{2}}{\mu_{3}}, \quad Re_E=\frac{\varepsilon_3 \dot{\gamma}_{0}}{\sigma_3},\\ &Ca_{e}=\frac{\varepsilon_{3}E^{2}R_{b}}{\gamma_{23}},\quad Wc=\frac{R_{b}}{H},\quad S_{ij}=\frac{\varepsilon_i}{\varepsilon_j},\quad R_{ij}=\frac{\sigma_{i}}{\sigma_{j}},
	\end{aligned}
	\label{eq51}
\end{equation}
where $Ca$ is the Capillary number, $Re$ is the Reynolds number, and $Re_E$ denotes the electric Reynolds number. $Ca_e$ represents the electric Capillary number.  $R_b$ corresponds to the initial radius of the droplet, and $\dot{\gamma}_0$ is the characteristic shear rate. Additionally, $Wc$ defines the confinement ratio, while $S_{ij}$ and $R_{ij}$ represent the ratios of permittivity and conductivity between phases $i$ and $j$, respectively. 

\subsection{Electro-Osmotic Flow in a Flat Microchannel}
\begin{figure}[htbp]
	\centering
	\includegraphics[width=0.4\textwidth]{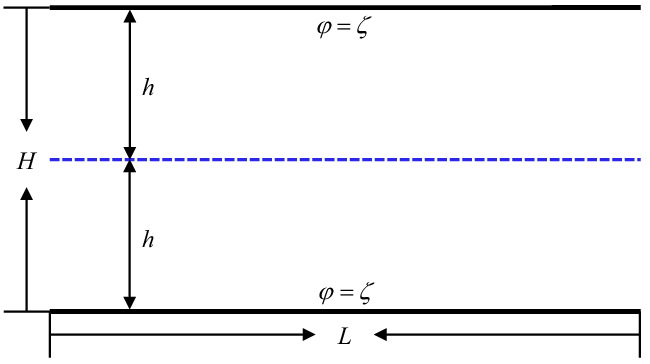}
	\caption{Schematic diagram of electro-osmotic flow in a parallel-plate microchannel.}
	\label{fig1}
\end{figure}

To verify our numerical framework, we first consider steady electro-osmotic flow (EOF) driven by an electric field, enabling comparison with analytical solutions \cite{Yoshida:CNSNS2014,Yang:JSC2014}. As shown in Fig. \ref{fig1}, the computational domain is a straight rectangular microchannel of length $L_x = 5.0\ \mu\text{m}$ and height $H = 2h = 1.0\ \mu\text{m}$, filled with a symmetric binary electrolyte like aqueous NaCl or KCl. To complement the hydrodynamic field and electric potential governed by the NS and Poisson equations [see Eqs. \eqref{eq27a}, \eqref{eq27b}, and \eqref{eq27d}], the evolution of the $i$-th ionic component ($i = 1,2$) in the electrolyte solution is explicitly described. Introducing $n_i$ as the ionic number concentration, its transport is dictated by the Nernst–Planck equation \cite{Yang:JSC2014}:
\begin{equation}
	\frac{\partial n_i}{\partial t} + \mathbf{u} \cdot \nabla n_i = \nabla \cdot \left[ D_i \nabla n_i + \frac{z_i e D_i}{k_B T} n_i \nabla \varphi \right]
	\label{eq52}
\end{equation}
where  $\mathbf{u}$ is the fluid velocity vector, $D_i = 1.0 \times 10^{-8}~\text{m}^2/\text{s}$ denotes the diffusion coefficient, and $z_i$ represents the ionic valence. Furthermore, the physical constants are defined as the elementary charge  $e = 1.602 \times 10^{-19}~\text{C}$, the Boltzmann constant $k_B = 1.38 \times 10^{-23}~\text{J}/\text{K}$, and the absolute temperature $T = 273~\text{K}$, with $\varphi$ denoting the local electric potential.

Assuming a  low surface zeta potential ($|\zeta| \le$ 25 mV, satisfying $e\zeta / k_B T \ll 1$), the Debye-Hückel approximation becomes applicable, allowing for the linearization of the electrical double layer (EDL) charge interactions. For this specific regime, the analytical solutions for the transverse electrical potential $\varphi(y)$ and the axial velocity profile $u(y)$ are classically given as \cite{Qu:JCIS2000,Kamali:CP2018}:
\begin{equation}
	\begin{aligned}
		\varphi(y) = \zeta \frac{\cosh(\kappa y)}{\cosh(\kappa h)}, \quad \quad \kappa  = \frac{1}{\lambda_D} = \sqrt{\frac{e^2 \sum_i z_i^2 n_{i}}{\varepsilon k_B T}},\quad
		u(y) = -\frac{\varepsilon \zeta E}{\mu} \left( 1 - \frac{\cosh(\kappa y)}{\cosh(\kappa h)} \right)
	\end{aligned}
	\label{eq53}
\end{equation}	
where $\lambda_D$ is the characteristic Debye length, representing the thickness of the EDL.  The fluid permittivity  is $\varepsilon = 6.95 \times 10^{-20}~\text{C}^2/(\text{J}\cdot\text{m})$, the dynamic viscosity is $\mu = 1.0 \times 10^{-3}~\text{N}\cdot\text{s}/\text{m}^2$, and the strength of the  applied electric field is $E = 250~\text{V}/\text{m}$. Additionally, the fluid density is set to $\rho = 1.0 \times 10^3~\text{kg}/\text{m}^3$.

\begin{figure}[htbp]
	\centering
	\subfigure[velocity]{\includegraphics[width=0.415\textwidth]{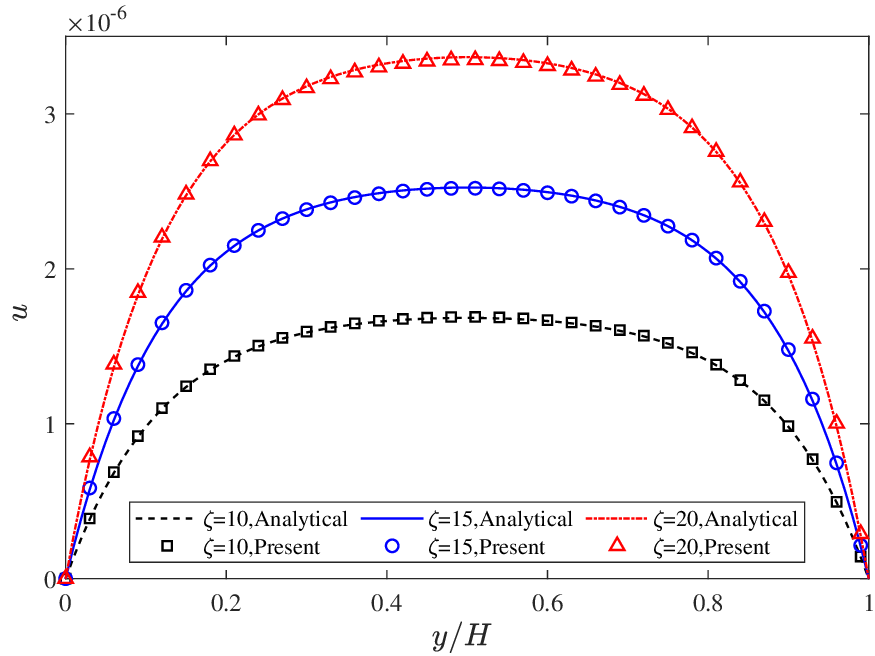}}\quad\quad
	\subfigure[potential]{\includegraphics[width=0.4\textwidth]{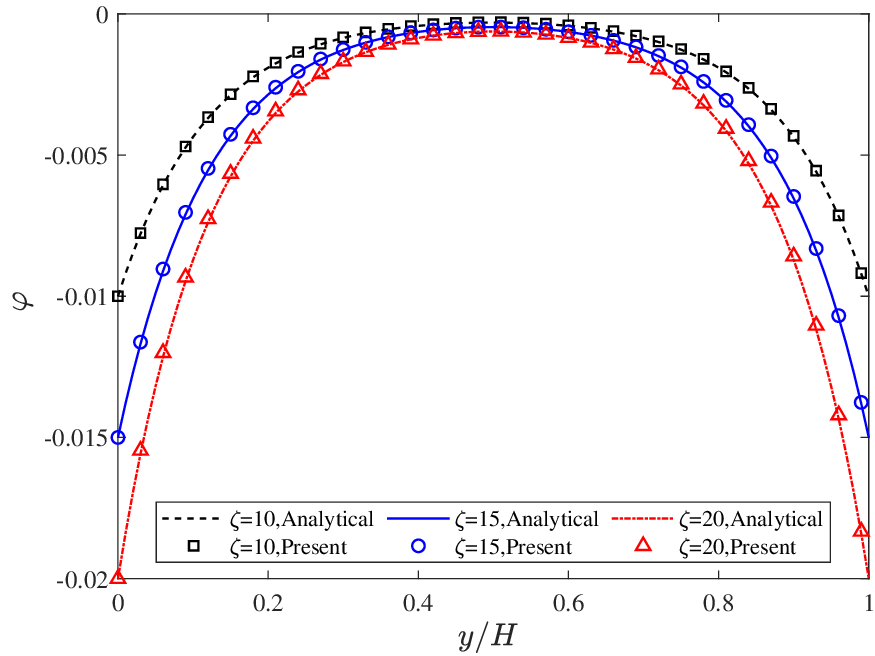}}
	\caption{The profiles along the $y$ coordinate for different $\zeta$ potentials: (a) velocity $u$; (b)  potential $\varphi$.}
	\label{fig2}
\end{figure}

In  this simulation, the computational domain is discretized with a grid resolution of $500 \times 100$. A comprehensive validation of the electric potential and velocity distributions along the $y$-direction is displayed in Figs. \ref{fig2}, demonstrating excellent agreement between the numerical predictions and analytical solutions across three varying Zeta potentials  ($\zeta = 10~\text{mV}$, $15~\text{mV}$, and $20~\text{mV}$).  This  correspondence verifies the capability of the current algorithm to accurately capture fundamental electrohydrodynamic transport phenomena, thereby validating its applicability.

\subsection{A static ternary-phase compound droplet}
To further validate the capability of the proposed LB method, we investigate a static ternary-phase compound droplet. The droplet configuration comprises an internal core ($\phi_1$) with a radius of $R_a = 50\delta x$, which is concentrically enveloped by an intermediate immiscible layer ($\phi_2$) with a radius of $R_b = 100\delta x$. This entire compound structure is  suspended within a surrounding bulk fluid phase ($\phi_3$).

\begin{figure}[htbp]
	\centering
	\subfigure[]{\includegraphics[width=0.32\textwidth]{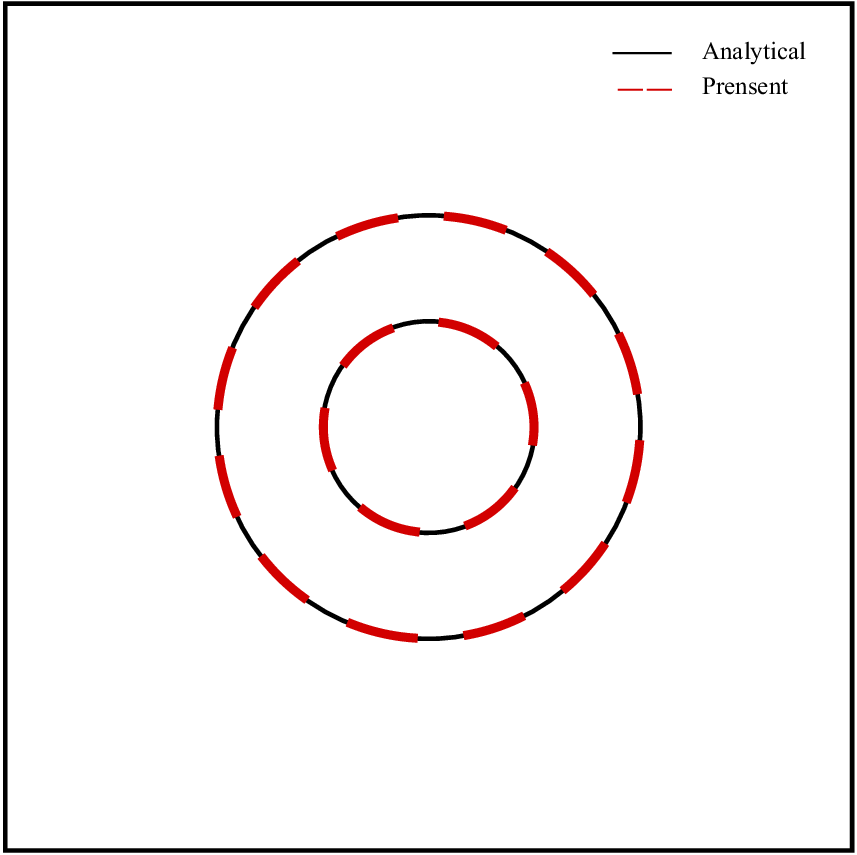}\label{fig3a}}
	\subfigure[]{\includegraphics[width=0.42\textwidth]{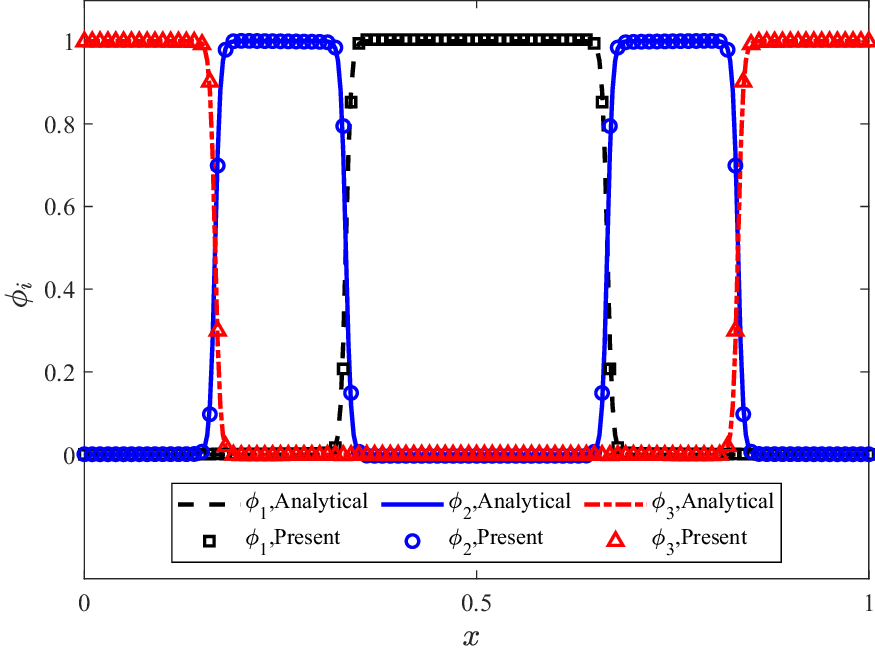}\label{fig3b}}
	\caption{(a)Equilibrium interfacial morphologies of the compound droplet: a comparison between numerical and analytical results; (b) Profile of the phase variable $\phi_i$ along the mid-plane defined by $y = H/2$.}
	\label{fig3}
\end{figure}
Initially, a stationary compound droplet is simulated to evaluate the accuracy of the proposed model in capturing complex multiphase interfaces. The physical properties are uniform across all three phases, specifically set to dynamic viscosities $\mu_1 = \mu_2 = \mu_3 =$ 0.1, interfacial tensions $\gamma_{ij} =$ 0.001 (for $i \neq j$), and fluid densities $\rho_1 = \rho_2 = \rho_3 =$ 1.  The initial volume fraction distributions for the three phases are prescribed as follows:
\begin{equation}
	\begin{aligned}
		\phi_1(x,y)&=\frac{1}{2}+\frac{1}{2}\tanh\frac{2[R_a-\sqrt{(x-N_x/2)^2+(y-N_y/2)^2}]}{W},\\
		\phi_2(x,y)&=(1-\phi_{1})\left\{\frac{1}{2}+\frac{1}{2}\tanh\frac{2[R_{b}-\sqrt{(x-N_x/2)^2+(y-N_y/2)^2}]}{W}\right\}.\\
		\phi_3(x,y)&=1-\phi_1(x,y)-\phi_2(x,y);
	\end{aligned}
	\label{eq54}
\end{equation}

Fig. \ref{fig3a} shows excellent agreement between the predicted terminal equilibrium morphology of the compound droplet and its analytical solution, confirming the high accuracy of our LB method for multiphase interface tracking. This is further corroborated by the volume fraction $\phi_i$ profiles along the axial centerline in Fig. \ref{fig3b}.

After validating accuracy, we quantitatively evaluate the convergence rate by computing global relative errors GREs for volume fractions at different grid resolutions. For any variable 
$\varphi$, the GRE is defined as:
\begin{equation}
	\text{GRE}_{\varphi} = \frac{\sum_{i,j} |\varphi_{\text{num}} - \varphi_{\text{ana}}|}{\sum_{i,j} |\varphi_{\text{ana}}|}
	\label{eq55}	
\end{equation}
where the subscripts "num" and "ana" denote the numerical outcomes and analytical solutions, respectively. As detailed in the Table  \ref{table1}, the evaluated GREs exhibit a consistent reduction as the mesh is refined. Specifically, the spatial convergence rate is quantified using the formula $\text{rate} = \log_2(E_{2h} / E_h)$, where $E_{2h}$ and $E_h$ denote the GREs corresponding to the mesh sizes $2h$ and $h$, respectively. This rigorous error analysis demonstrates that the current numerical framework successfully achieves a second-order convergence rate in space.
\begin{table}[htbp]
	\caption{Global relative errors and spatial convergence rates for $\phi_i$ across different mesh resolutions.}
	\centering
	\setlength{\tabcolsep}{4mm}
	\resizebox{1.0\textwidth}{!}{
	\begin{tabular}{lllllll}
		\specialrule{0.05em}{1pt}{1pt}\specialrule{0.05em}{1pt}{1pt}
		\makebox[0.03\textwidth][l]{} &Err$(\phi_1)$ &rate$(\phi_1)$    &Err$(\phi_2)$ &rate$(\phi_2)$   &Err$(\phi_3)$ &rate$(\phi_3)$     \\ \hline\specialrule{0em}{1pt}{1pt}
		$\delta x=1/128$ &$8.42\times10^{-2}$                 & - &$1.08\times10^{-1}$  & -  &$5.04\times10^{-2}$  & -   \\
		$\delta x=1/256$ 	&$2.76\times10^{-2}$     &1.61 &$3.40\times10^{-2}$  &1.67  &$1.18\times10^{-2}$  &2.10    \\
		$\delta x=1/512$ 	&$8.99\times10^{-2}$      &1.62 &$1.03\times10^{-3}$  &1.73  &$3.28\times10^{-3}$  &1.85    \\ 
		\specialrule{0.05em}{1pt}{1pt}\specialrule{0.05em}{1pt}{1pt} 
	\end{tabular}}
	\label{table1}
\end{table}

\subsection{Spreading of a liquid lens}

\begin{figure}[htbp]
	\centering
	\includegraphics[width=0.25\textwidth]{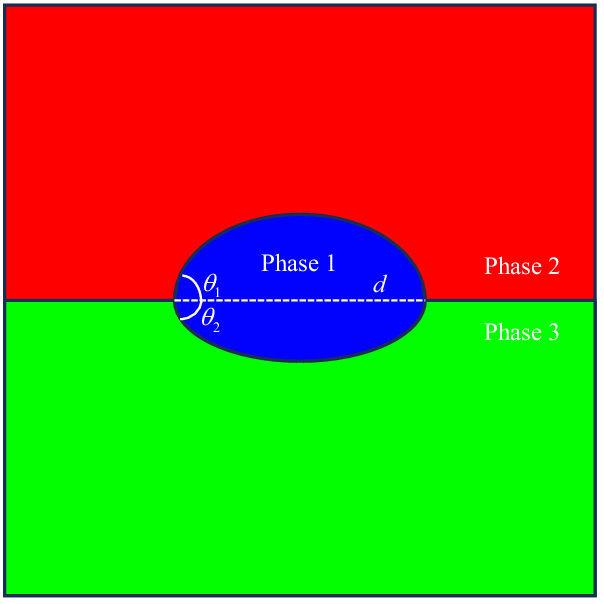}
	\caption{Diagram of the equilibrium lens morphology.}
	\label{fig4}
\end{figure}

The classic benchmark of a spreading liquid lens is investigated here to assess the capability of the present LB model for three-phase flows. The domain uses $N_x \times N_y = 200 \times 200$ grid, periodic boundaries in $x$, and no-slip bounce-back on top and bottom walls. Parameters: densities $\rho_1 = 10$, $\rho_2 = 1$, $\rho_3 = 5$; mobility $M_0 = 0.01$; interface thickness $D = 4.0$; relaxation times $\tau_1 = \tau_2 = \tau_g = 0.8$. Initially, as shown in Fig.~\ref{fig4}, a circular lens of radius $R = 30$ is placed at $(x_c, y_c)$ between two immiscible fluid layers. The initial order parameters are set as:
\begin{equation}
	\begin{aligned}
		\phi_{1}(x,y)&=0.5+0.5\tanh\left[\frac{2\left(R-\sqrt{(x-x_{c})^{2}+(y-y_{c})^{2}}\right)}{D}\right],\\
		\phi_{2}(x,y)&=\max\left[0.5+0.5\tanh\frac{2(y-y_{c})}{D}-\phi_{1}(x,y),0\right]\\
		\phi_{3}(x,y)&=1-\phi_{1}(x,y)-\phi_{2}(x,y)
	\end{aligned}
	\label{eq56}
\end{equation}

Driven by interfacial forces, the initially circular lens deforms to thermodynamic equilibrium. Neumann's law determines the final contact angles $\theta_1$, $\theta_2$ from surface tensions \cite{Kim:CCP2012}:

\begin{equation}
	\cos(\theta_1)=\frac{\gamma_{23}^2+\gamma_{12}^2-\gamma_{13}^2}{2\gamma_{23}\gamma_{12}},\quad\cos(\theta_2)=\frac{\gamma_{23}^2+\gamma_{13}^2-\gamma_{12}^2}{2\gamma_{23}\gamma_{13}}
	\label{eq57}
\end{equation}

Furthermore, based on the lens area $A$, the theoretical distance ($d$) between the two triple junctions can be derived as \cite{Kim:CCP2012}
\begin{equation}\left(\frac{d}{2}\right)^2\sum_{j=1}^2\frac{1}{\sin(\theta_j)}\left[\frac{\theta_j}{\sin(\theta_j)}-\cos(\theta_j)\right]=A
	\label{eq58}
\end{equation}

\begin{figure}[htbp]
	\centering
	\subfigure[]{\includegraphics[width=0.25\textwidth]{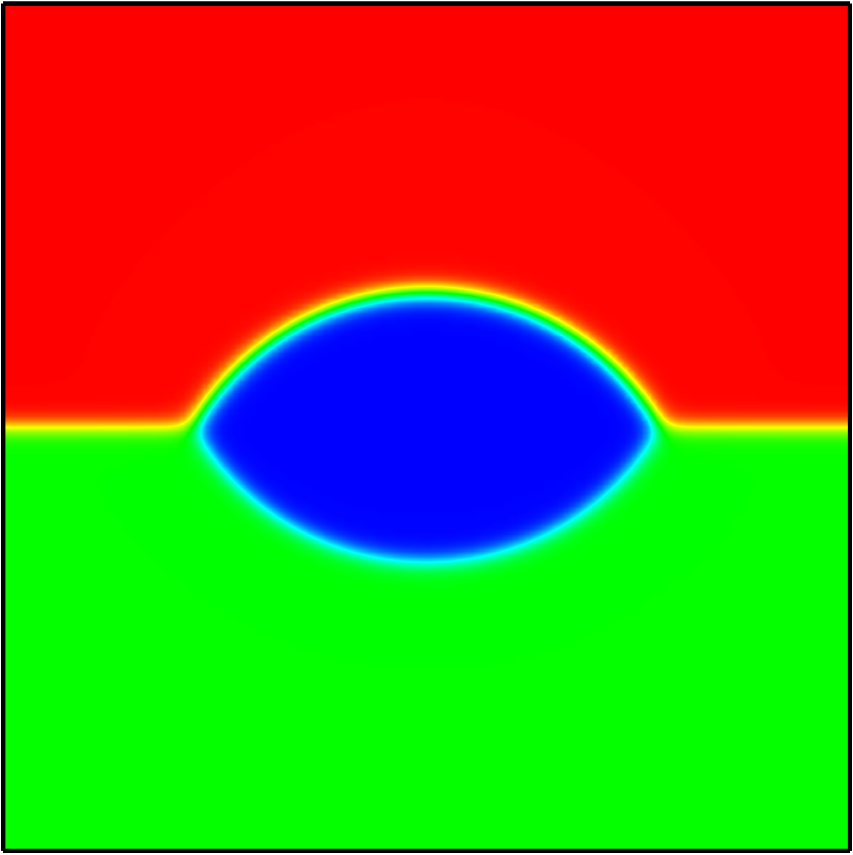}}\quad\quad
	\subfigure[]{\includegraphics[width=0.25\textwidth]{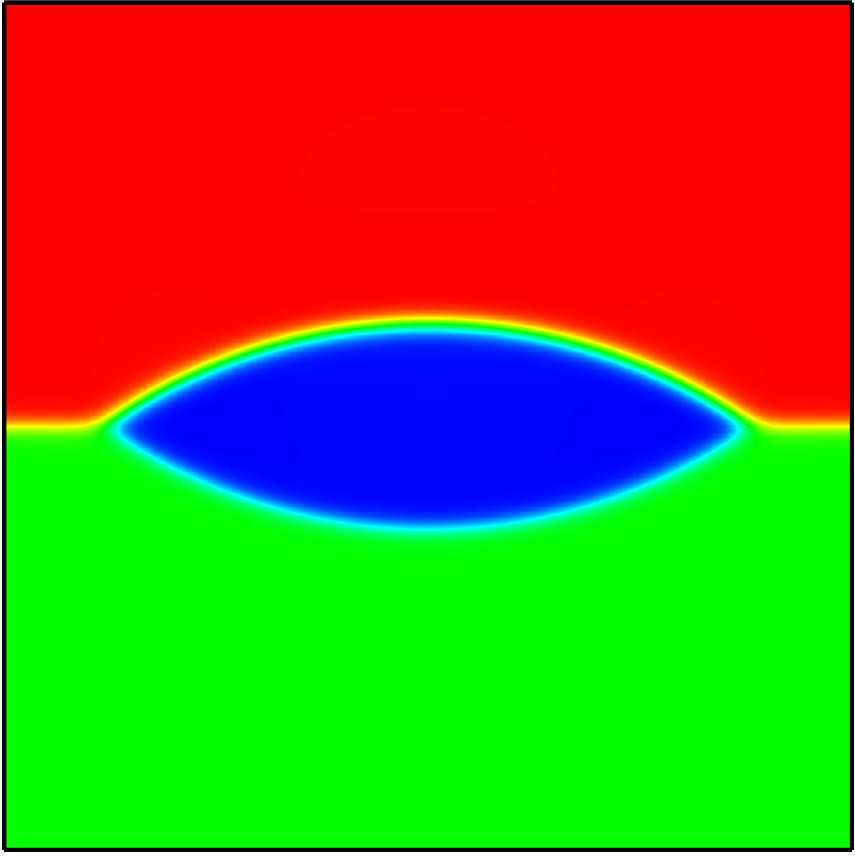}}\quad\quad
	\subfigure[]{\includegraphics[width=0.25\textwidth]{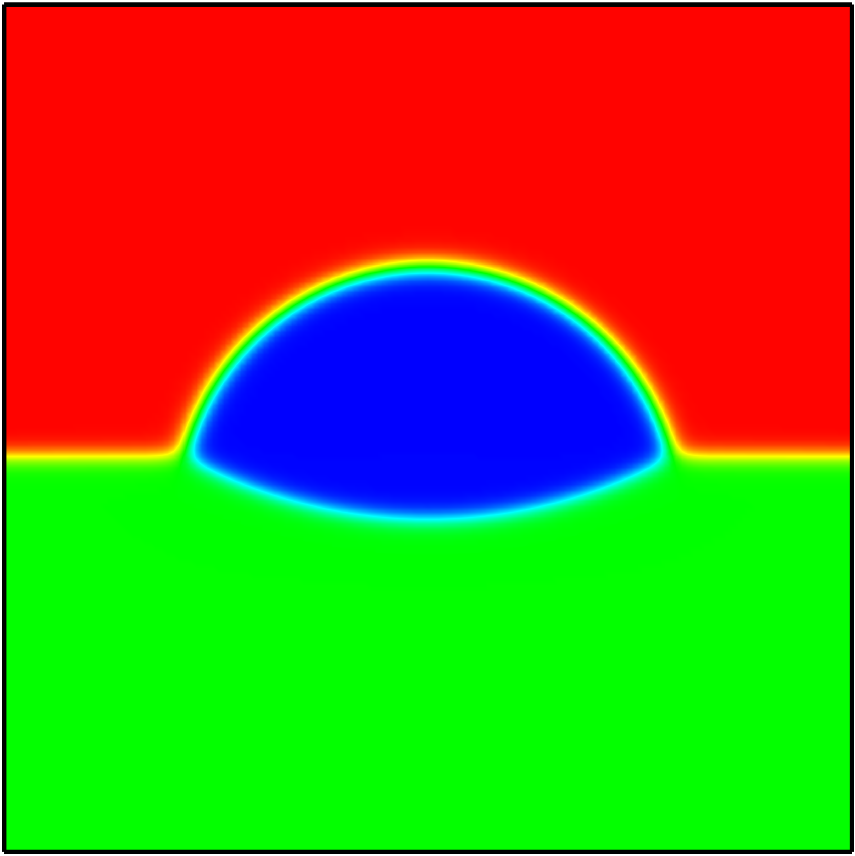}}
	\caption{The equilibrium states of  lens with different surface tension ratios: (a) $\gamma_{12}:\gamma_{13}:\gamma_{23} = 1:1:1$ ; (b) $\gamma_{12}:\gamma_{13}:\gamma_{23} = 0.6:0.6:1$ ; (c) $\gamma_{12}:\gamma_{13}:\gamma_{23} = 1:2:2$.}
	\label{fig5}
\end{figure}

In our simulations, three distinct surface tension ratios ($\gamma_{12}:\gamma_{13}:\gamma_{23} = 1:1:1$, $0.6:0.6:1$, and $1:2:2$) are considered to induce different equilibrium shapes. As illustrated in Figs. \ref{fig5}, the liquid lens successfully evolves into distinct interfacial morphologies under varying capillary forces, aligning visually with established literature. For a rigorous quantitative assessment, the numerical predictions of the contact angles ($\theta_1, \theta_2$) and the spreading length ($d$) are extracted and compared against the theoretical solutions, as uniformly summarized in Table \ref{table2}. The comparison shows excellent agreement and low relative errors, confirming the model’s accuracy for three-phase interactions.
\begin{table}[htbp]
	\caption{Equilibrium Contact Angles ($\theta_1, \theta_2$) and Spreading Length ($d$) for Different Surface Tensions.}
	\centering
	\setlength{\tabcolsep}{1mm}
	\resizebox{1.0\textwidth}{!}{
		\begin{tabular}{lllllllllllll}
			\specialrule{0.05em}{1pt}{1pt}\specialrule{0.05em}{1pt}{1pt}
			\makebox[0.15\textwidth][l]{$\gamma_{12}:\gamma_{13}:\gamma_{23}$} &   &\multicolumn{3}{c}{Analytical solutions} &   \makebox[0.02\textwidth][l]{}  &\multicolumn{3}{c}{Present LB model}&   \makebox[0.02\textwidth][l]{} &\multicolumn{3}{c}{Relative errors}  \\ \specialrule{0em}{1pt}{0pt} \cmidrule(rr){3-5} \cmidrule(rr){7-9} \cmidrule(rr){11-13}
			& &\makebox[0.07\textwidth][c]{$d$} &\makebox[0.07\textwidth][c]{$\theta_1$ } & \makebox[0.07\textwidth][c]{$\theta_2$ }& &\makebox[0.07\textwidth][c]{$d$} &\makebox[0.07\textwidth][c]{$\theta_1$ } & \makebox[0.07\textwidth][c]{$\theta_2$ }&&\makebox[0.07\textwidth][c]{$d$} &\makebox[0.07\textwidth][c]{$\theta_1$ } & \makebox[0.07\textwidth][c]{$\theta_2$ } \\	\hline
			$1:1:1$  &    &83.12  &${60}^\circ$ &${60}^\circ$& &81.2 &${61.2}^\circ$ &${61.2}^\circ$& &$2.30\%$  &$2.00\%$ &$2.00\%$\\
			
			$0.6:0.6:1$     &  &117.56 &${33.56}^\circ$ &${33.56}^\circ$& &114.22  &${34.50}^\circ$ &${34.50}^\circ$& &$2.84\%$  &$2.80\%$ &$2.80\%$              \\
			
			$1:2:2$    &  &86.94 &${75.52}^\circ$ &${28.96}^\circ$&  &85.13  &${75.9}^\circ$ &${30.7}^\circ$& &$2.08\%$ &$0.50\%$ &$6.00\%$                 \\
			
			\specialrule{0.05em}{1pt}{1pt}\specialrule{0.05em}{1pt}{1pt}
	\end{tabular}}
	\label{table2}
\end{table}
\subsection{Three-phase compound droplet deformation}
To systematically evaluate the electrohydrodynamic behavior of the ternary-phase compound droplet, a series of numerical simulations are conducted. The extent of morphological change is quantitatively characterized by the deformation factor, defined as $D_{ij} = (r_{ij,\parallel} - r_{ij,\perp}) / (r_{ij,\parallel} + r_{ij,\perp})$, wherein $r_{ij,\parallel}$ and $r_{ij,\perp}$ denote the droplet radii parallel and perpendicular to the applied uniform electric field, respectively. Under the theoretical assumption of small deformations, the analytical solutions for the inner ($D_{12}$) and outer ($D_{23}$) deformation factors can be explicitly expressed as \cite{Behjatian:SMP2013,Behjatian:AM2015}:

\begin{equation}	
	D_{12} = \frac{81}{16} \frac{S_{23}\lambda}{\Omega} \Gamma^2 \Phi_{12} Ca_E, \quad D_{23} = \frac{9}{16} S_{23} \Gamma^2 \Phi_{23} Ca_E,
	\label{eq59}
\end{equation}
where $\Phi_{12/23} = \Phi_{12/23}^e + \Phi_{12/23}^h$ act as the characteristic functions determining the sense of the inner and outer droplet deformations. The parameters $\Phi_{12/23}^e$ and $\Phi_{12/23}^h$ denote the respective components of the net normal Maxwell electric stress and hydrodynamic stress, which are explicitly expanded in the following form:
\begin{equation}
	\begin{aligned}
		\Phi_{12}^e &= R_{12}^2 + 1 - 2S_{12},\\
		\Phi_{12}^h &= -\left[ \frac{3(3\lambda^5 \Lambda_E K_{13} + (6\lambda^2 \Lambda_A + 8\Lambda_B - 2\lambda^5 \Lambda_C + \lambda^7 \Lambda_D) K_{23})}{\Pi} \right] (S_{12} - R_{12}),\\
		\Phi_{23}^e &= \frac{(R_{12}+2)^2 (1-\zeta)^2}{S_{23}} \left[ \left(\frac{1+2\zeta}{1-\zeta}\right)^2 R_{23}^2 + 1 - \left(\left(\frac{1+2\zeta}{1-\zeta}\right)^2 + 1\right) S_{23} \right],\\
		\Phi_{23}^h &= -\left[ \frac{27\lambda^5 (2\Lambda_H - (6\Lambda_A + 8\Lambda_B - 2\Lambda_C + \Lambda_D) K_{23})}{\Pi} \right] (S_{12} - R_{12}).
	\end{aligned}
	\label{eq60}
\end{equation}
It should be noted that the detailed expressions for the associated parameter coefficients $\Gamma$, $\zeta$, $\Pi$, and $\Lambda_{A-H}$ are systematically documented and can be found in Ref. \cite{Behjatian:SMP2013}. Fundamentally, the interfacial dynamics are driven by the induced electrohydrodynamic stresses, which are intrinsically linked to the free surface charge distribution. The local sign of the free charge $q_{ij}$ at the phase boundaries is  governed by the relative dominance of the physical property ratios $R_{ij}/S_{ij}$, analytically given by \cite{Behjatian:SMP2013}:
\begin{equation}
	\begin{aligned}
		\frac{q_{12}}{\epsilon_3 E_\infty}& = 9\Gamma S_{23}S_{12} (1 - R_{12}/S_{12}) \cos\theta,\\
		\frac{q_{23}}{\epsilon_3 E_\infty} &= 3\Gamma S_{23} (R_{12} + 2)(1 + 2\zeta) (1 - R_{23}/S_{23}) \cos\theta.
	\end{aligned}
	\label{eq61}
\end{equation}

\begin{figure}[htbp]
	\centering
	\includegraphics[width=0.28\textwidth]{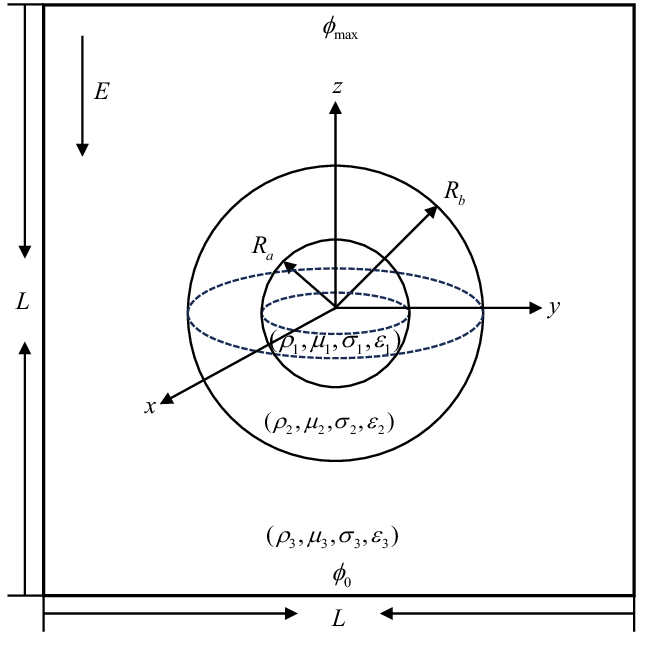}
	\caption{Physical model of a three-phase compound droplet in a uniform electric field.}
	\label{fig6}
\end{figure}

As depicted in Fig. \ref{fig6}, in a three-dimensional space, the droplet radii satisfy $R_b = 2R_a = 100\delta x$. In the following simulations, to effectively stabilize the compound droplet and neglect buoyancy effects, several baseline parameters are strictly fixed: the interfacial tensions $\gamma_{ij} = 0.001\ (i \neq j)$, the dynamic viscosities $\mu_1 = \mu_2 = \mu_3 = 0.1$, and the fluid densities $\rho_1 = \rho_2 = \rho_3 = 1$.

\begin{figure}[htbp]
	\centering
	\subfigure[]{\includegraphics[width=0.36\textwidth]{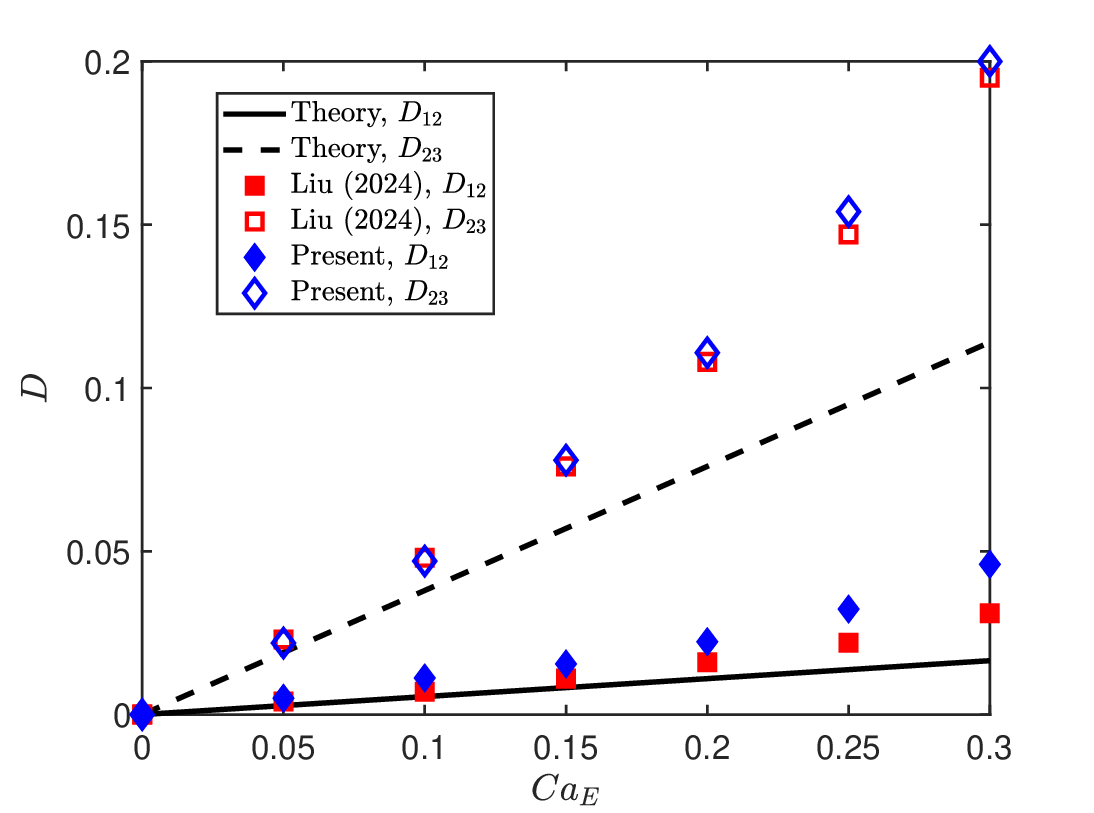}\label{fig7a}}
	\subfigure[]{\includegraphics[width=0.25\textwidth]{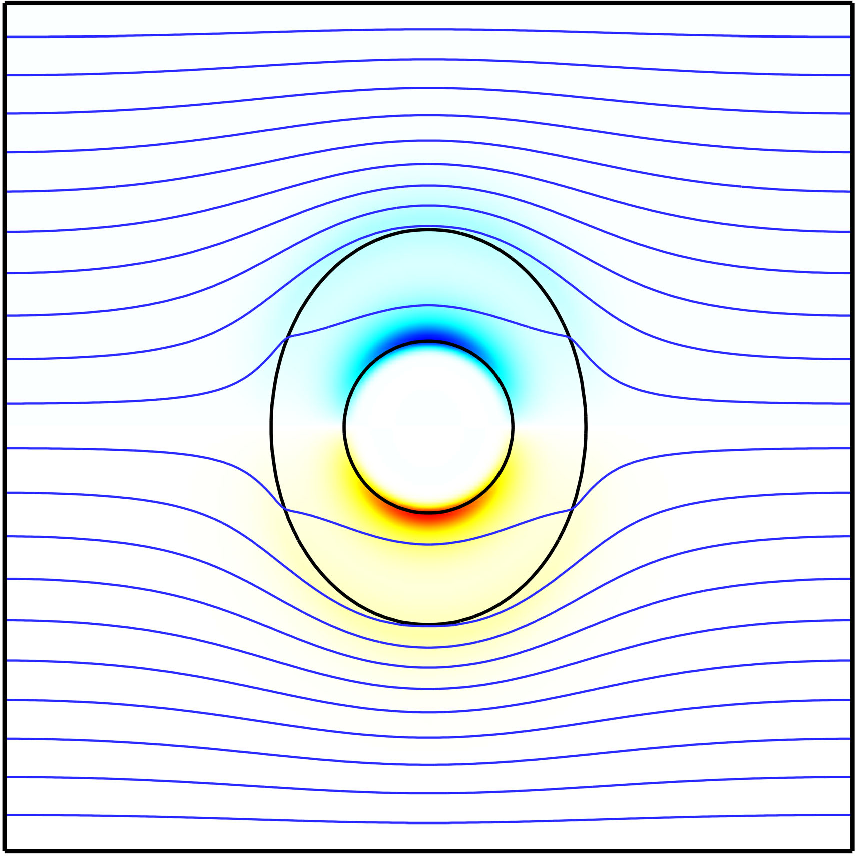}\label{fig7b}}\quad
	\subfigure[]{\includegraphics[width=0.25\textwidth]{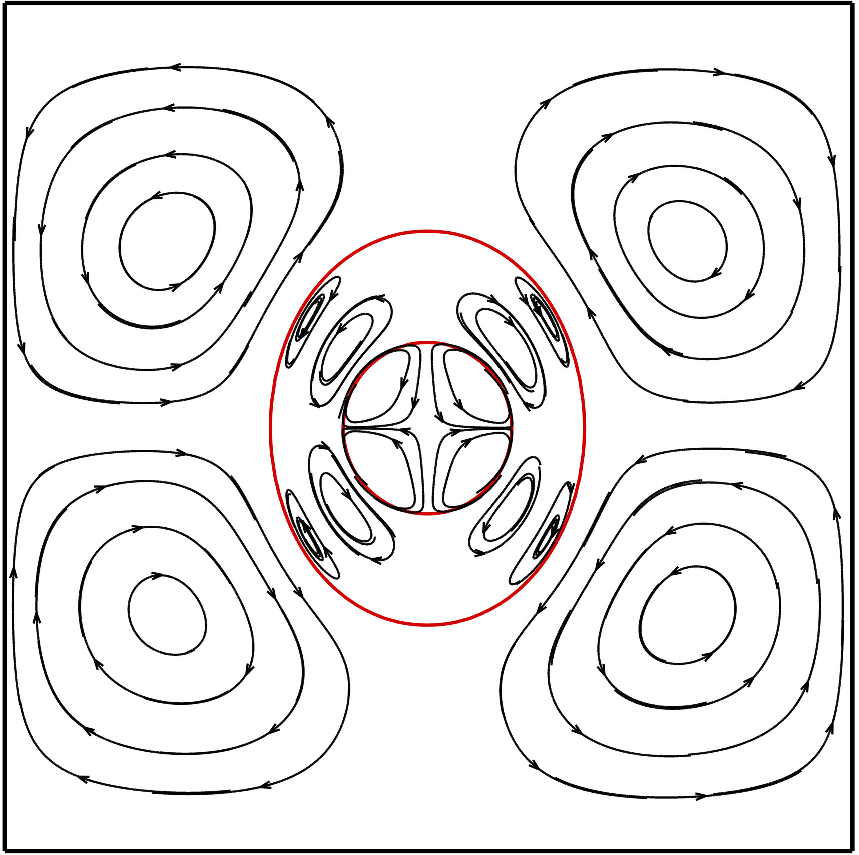}\label{fig7c}}
	\caption{For a compound droplet with parameters $(R_{12}, R_{23}, S_{12}, S_{23}) = (10, 8, 0.1, 5)$ in  electric field: (a) effect of electric strength $Ca_E$ on deformation; (b) free charge and electric potential at $Ca_E = 0.2$; (c) deformation and flow pattern at  $Ca_E = 0.2$.}
	\label{fig7}
\end{figure}

Initially, we first investigate a representative regime with the physical parameters set to $(R_{12}, R_{23}, S_{12}, S_{23}) = (10, 8, 0.1, 5)$. In this case, both the internal core and the external shell undergo longitudinal elongation parallel to the electric field direction ($D_{12} > 0, D_{23} > 0$). As illustrated in Fig. \ref{fig7a}, the magnitude of prolate deformation for both interfaces intensifies monotonically with increasing $Ca_E$, maintaining excellent agreement with the analytical solutions within the small deformation limit.To elucidate the underlying mechanisms, the corresponding physical fields are presented in Fig. \ref{fig7b}. Regarding the electric potential, the conductivity exhibits a monotonic decline from the innermost core towards the bulk fluid ($R_{12} > R_{23} > 1$). Due to the significantly higher electrical conductivity of the internal core compared to the surrounding layers, the electric field is effectively shielded, resulting in a near-zero field strength within the core. Furthermore, since the condition $R_{ij}/S_{ij} > 1$ is satisfied across both interfaces, negative free charges predominantly accumulate on the upper hemispheres of the droplet surfaces, while the lower hemispheres are occupied by positive charges, consistent with the theoretical predictions of Eq. (\ref{eq61}). The resulting electrohydrodynamic flow patterns are shown in Fig. \ref{fig7c}. The two counter-rotating vortices are formed within the intermediate shell. The fluid in the bulk phase surrounding the outer droplet moves from the equator toward the poles, while the fluid within the innermost core flows in the opposite direction from the poles toward the equator.

\begin{figure}[htbp]
	\centering
	\subfigure[]{\includegraphics[width=0.36\textwidth]{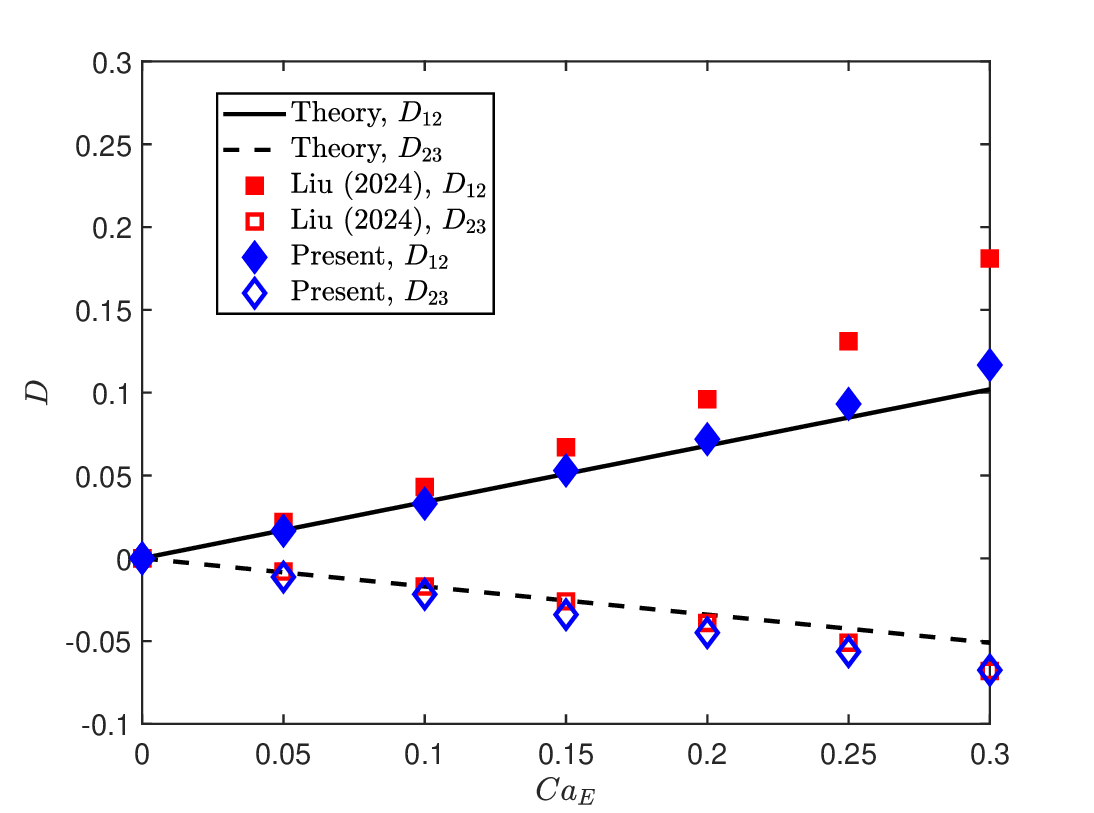}\label{fig8a}}
	\subfigure[]{\includegraphics[width=0.25\textwidth]{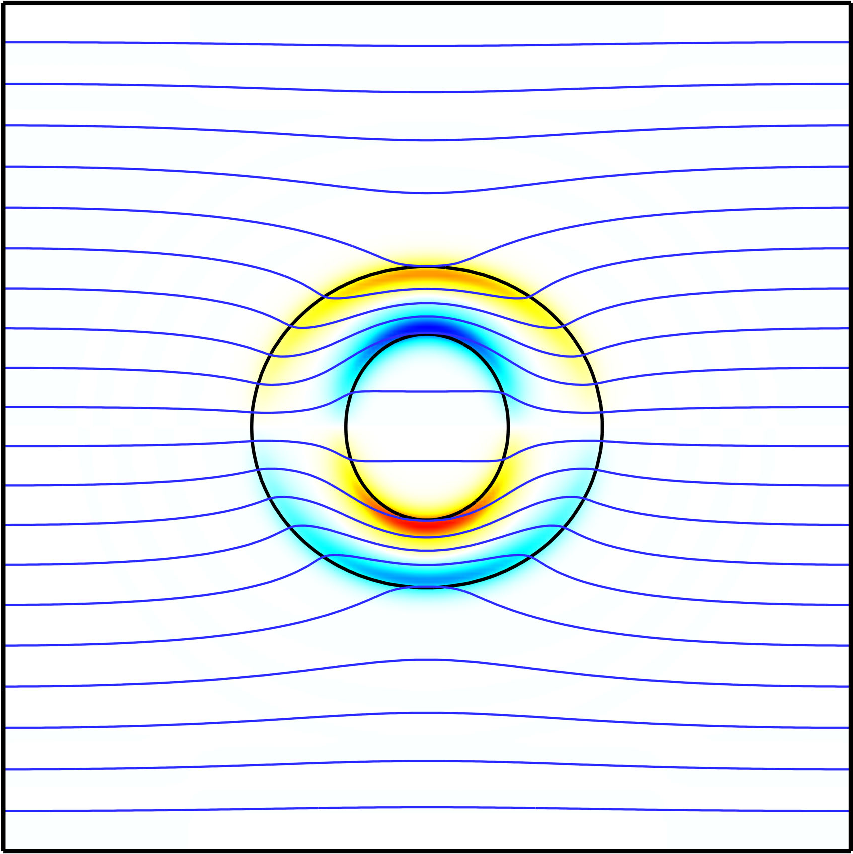}\label{fig8b}}\quad
	\subfigure[]{\includegraphics[width=0.25\textwidth]{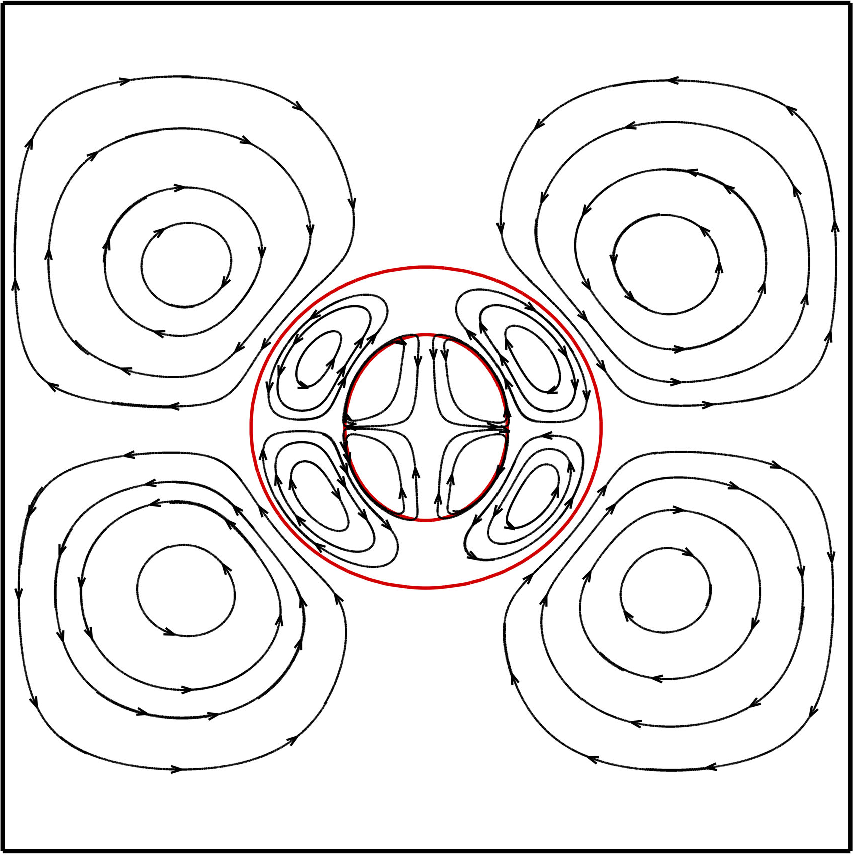}\label{fig8c}}
	\caption{For a compound droplet with parameters $(R_{12}, R_{23}, S_{12}, S_{23}) = (10, 0.1, 2, 0.5)$ in  electric field: (a) effect of electric strength $Ca_E$ on deformation; (b) free charge and electric potential at $Ca_E = 0.2$; (c) deformation and flow pattern at  $Ca_E = 0.2$.}
	\label{fig8}
\end{figure}

Subsequently, we investigate a divergent deformation pattern wherein the inner droplet elongates longitudinally ($D_{12} > 0$) while the outer droplet compresses equatorially ($D_{23} < 0$). This specific morphological state is achieved by configuring the parameters as $(R_{12}, R_{23}, S_{12}, S_{23}) = (10, 0.1, 2, 0.5)$. As depicted in Fig. \ref{fig8a}, the absolute values of both deformation factors amplify proportionally as $Ca_E$ scales up. Fig. \ref{fig8b} illustrates the steady-state electric field distribution for this scenario. In this configuration, the intermediate shell layer exhibits a notably lower electrical conductivity compared to both the central core and the ambient bulk fluid ($R_{12} > 1 > R_{23}$).  The electric potential lines become densely compacted within the intermediate shell, indicating a significant localized concentration of the electric field strength. Driven by these localized electric forces, the fluid flow structures are fundamentally altered compared to the first case. For this specific configuration ($R_{12}/S_{12} > 1$ and $R_{23}/S_{23} < 1$), Fig. \ref{fig8c} reveals that only a single vortex is present within the intermediate shell. The external flow is directed from the poles toward the equator. While this same direction is followed by the flow at the internal surface of the inner droplet, the fluid moves in the opposite direction within the intermediate shell.
\begin{figure}[htbp]
	\centering
	\subfigure[]{\includegraphics[width=0.36\textwidth]{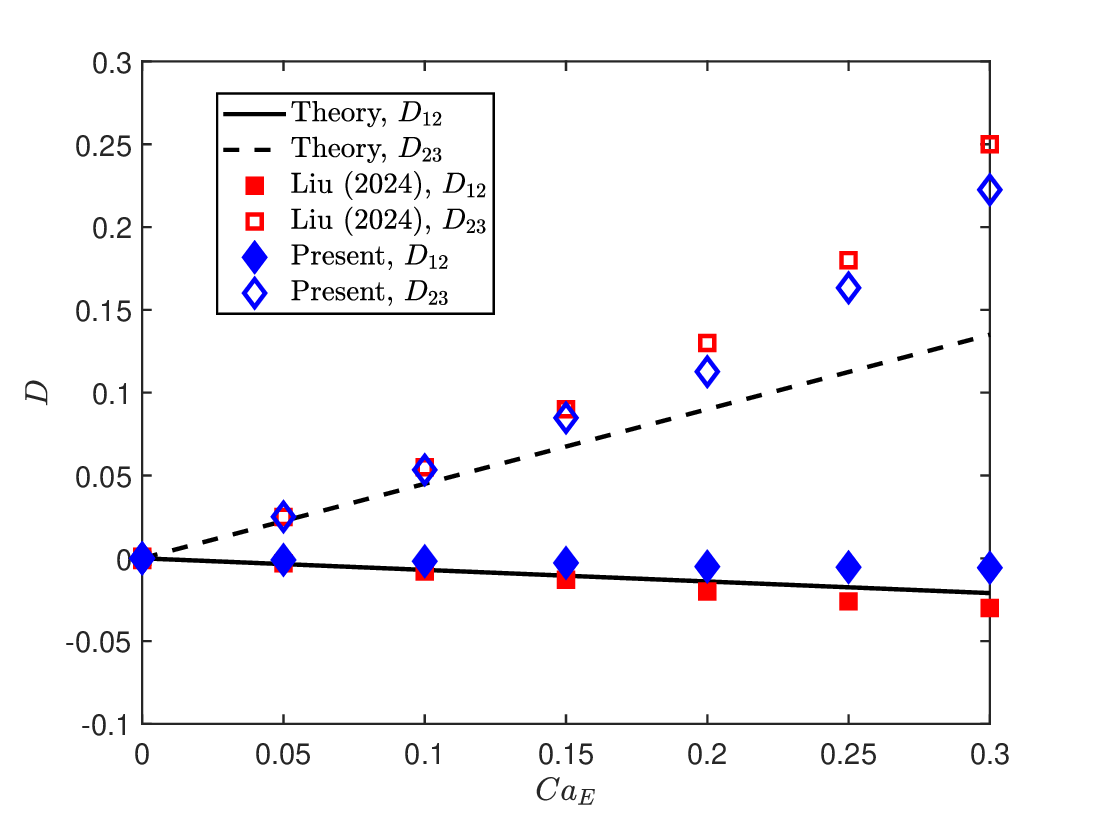}\label{fig9a}}
	\subfigure[]{\includegraphics[width=0.25\textwidth]{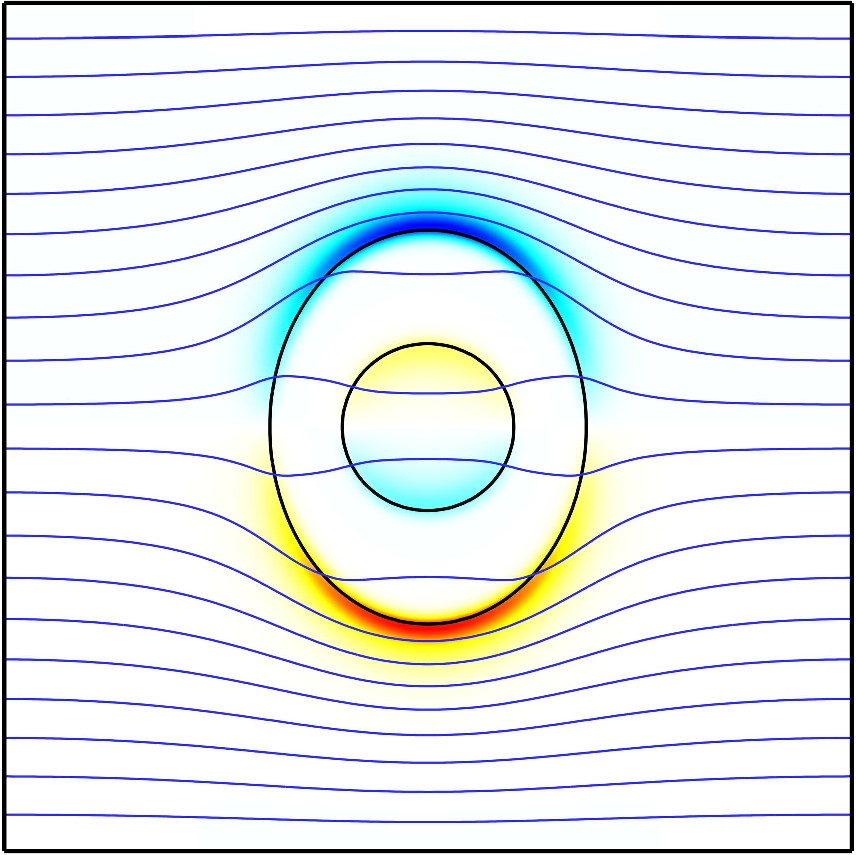}\label{fig9b}}\quad
	\subfigure[]{\includegraphics[width=0.25\textwidth]{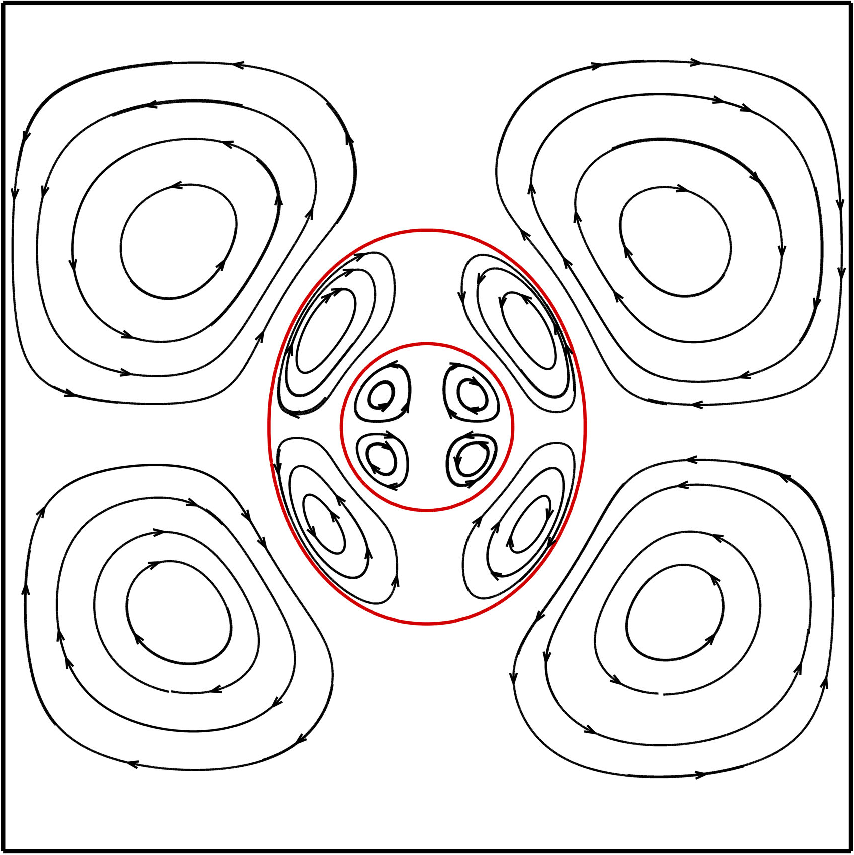}\label{fig9c}}
	\caption{For a compound droplet with parameters $(R_{12}, R_{23}, S_{12}, S_{23}) =(0.1, 10, 2, 0.5)$ in  electric field: (a) effect of electric strength $Ca_E$ on deformation; (b) free charge and electric potential at $Ca_E = 0.2$; (c) deformation and flow pattern at  $Ca_E = 0.2$ }
	\label{fig9}
\end{figure}

Finally, the third deformation pattern is characterized by an equatorially compressed inner core ($D_{12} < 0$) concentrically enveloped by a longitudinally elongated outer shell ($D_{23} > 0$). The parameter set defining this regime is $(R_{12}, R_{23}, S_{12}, S_{23}) = (0.1, 10, 2, 0.5)$. Consistent with the behavioral trends of the previous cases, Fig. \ref{fig9a} demonstrates that a continuous increment in $Ca_E$ robustly drives the divergent deformation of the two fluid interfaces. As visually confirmed in Fig. \ref{fig9b} , the electric potential lines within the shell are sparsely distributed, indicating a relatively weak electric field intensity. This behavior arises because the electrical conductivity of the intermediate shell is substantially higher than that of both the inner core and the surrounding bulk fluid, thereby reducing the local electric field strength in the highly conductive region. For this regime ($R_{12}/S_{12} < 1$ and $R_{23}/S_{23} > 1$), As shown in Fig. \ref{fig9c}, the flow characteristics in the bulk phase surrounding the outer droplet are similar to those in the first case, moving from the equator toward the poles. However, a single-vortex structure is present within the intermediate shell. The flow direction within the innermost core is from the poles toward the equator, which is opposite to the flow direction in the external bulk fluid.

In summary for the three typical deformation regimes presented in Figs.  \ref{fig7} to \ref{fig9}, the current numerical predictions agree excellently with the analytical solutions in the small deformation limit. Furthermore, the results highly align with the established findings in Ref. \cite{Liu:PD2024} regarding interfacial morphologies and flow fields. These comparisons fully verify the accuracy and robustness of the proposed numerical framework in simulating complex  multiphase EHD flows.

\begin{figure}[htbp]
	\centering
	\subfigure[$t=0$]{\includegraphics[width=0.15\textwidth]{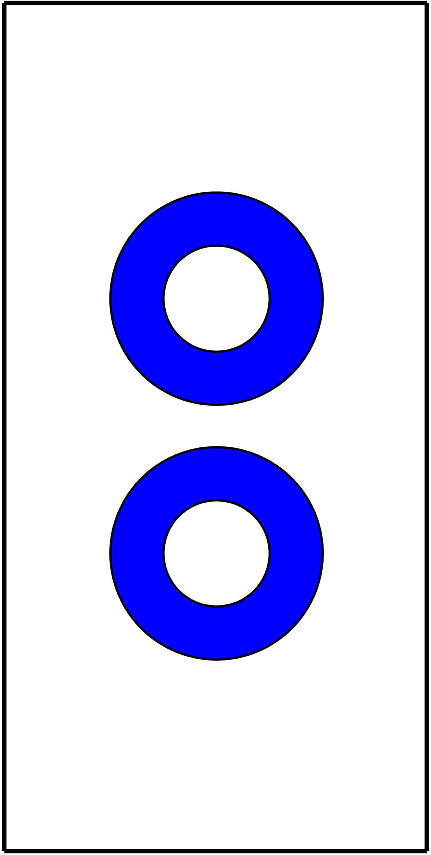}}\quad
	\subfigure[$t=2.5$]{\includegraphics[width=0.15\textwidth]{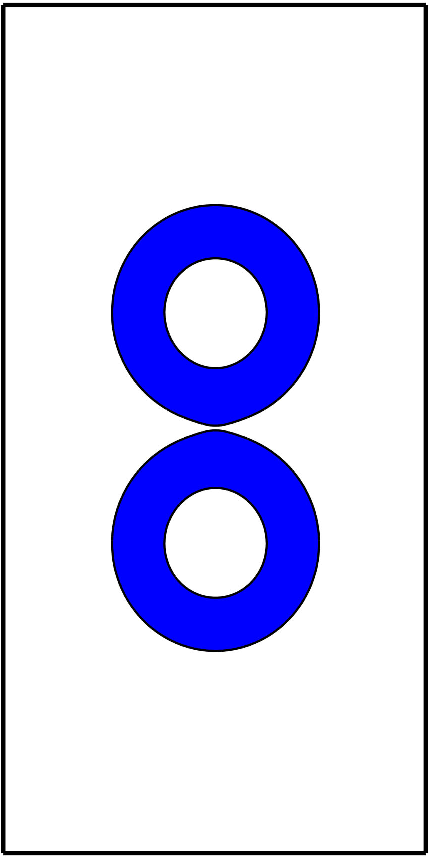}}\quad
	\subfigure[$t=5.0$]{\includegraphics[width=0.15\textwidth]{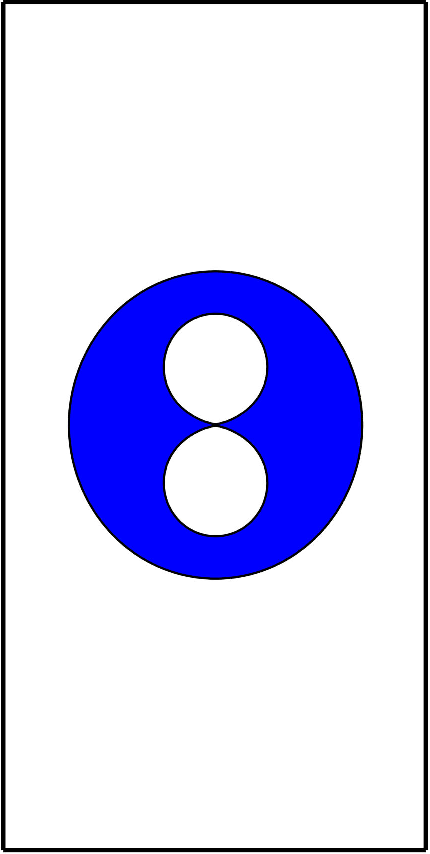}}\quad
	\subfigure[$t=8.0$]{\includegraphics[width=0.15\textwidth]{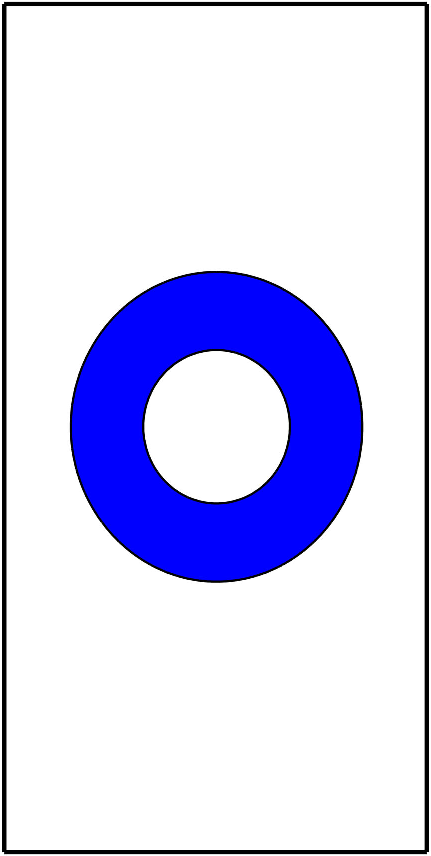}}
	\caption{Time evolution of the coalescence of compound droplets.}
	\label{fig10}
\end{figure}

\begin{figure}[htbp]
	\centering
	\subfigure[$t=0$]{\includegraphics[width=0.152\textwidth]{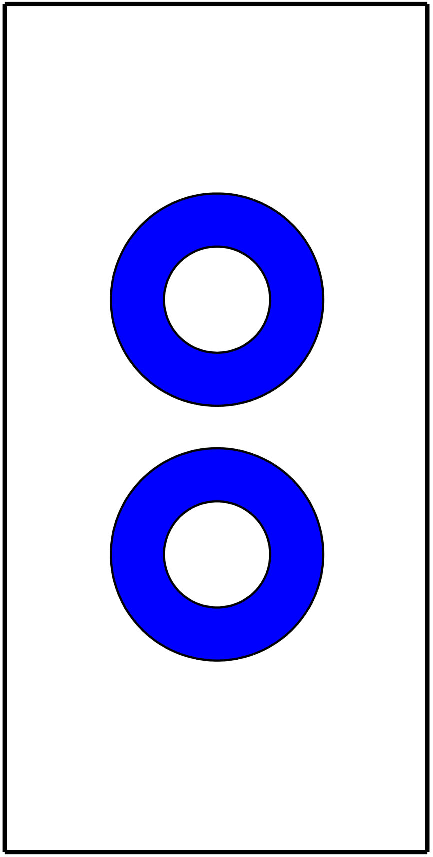}}\quad
	\subfigure[$t=50$]{\includegraphics[width=0.15\textwidth]{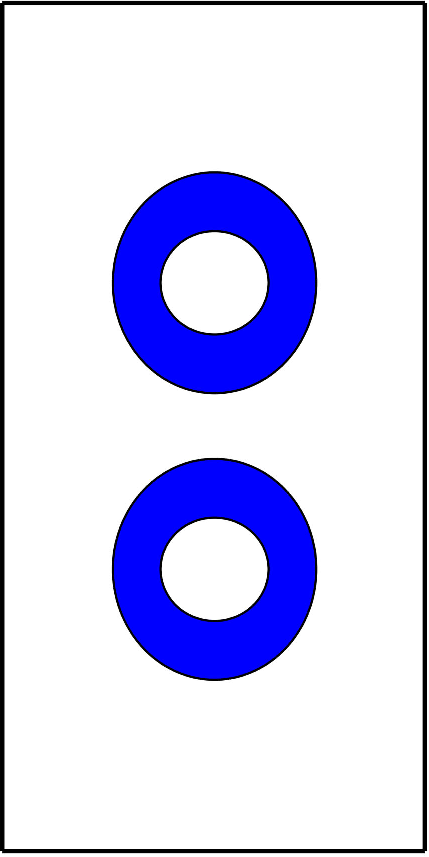}}\quad
	\subfigure[$t=100$]{\includegraphics[width=0.153\textwidth]{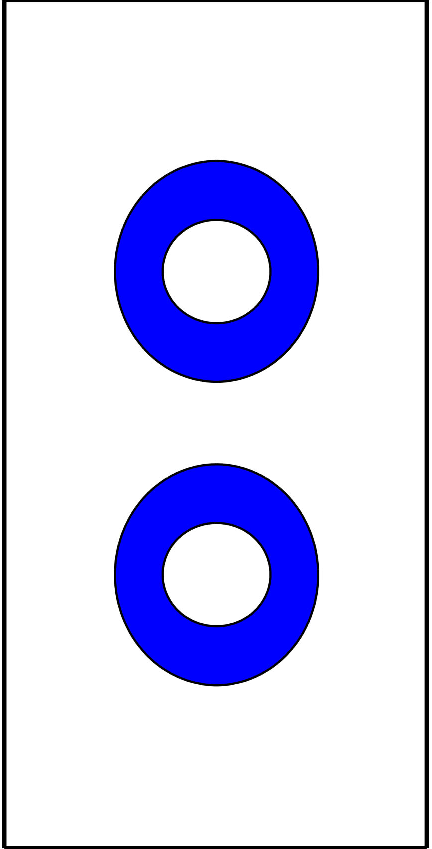}}\quad
	\subfigure[$t=500$]{\includegraphics[width=0.15\textwidth]{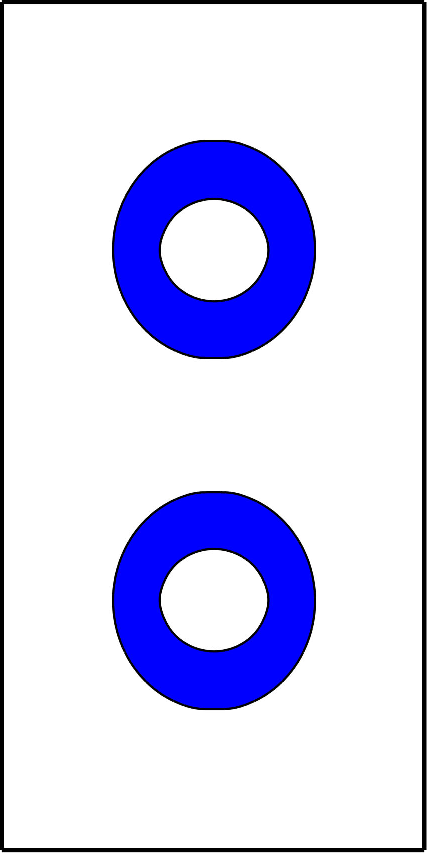}}
	\caption{Time evolution of the separation of compound droplets.}
	\label{fig11}
\end{figure}

\subsection{Electrohydrodynamic  coalescence and  separation of two compound droplets}

This section studies electrohydrodynamic interaction of two three-phase compound droplets in an immiscible matrix under a uniform electric field. The inner and outer radii are $R_{a}$ and $R_b$; the droplets are initially symmetric along the field direction. The electric capillary number is fixed at $Ca_E=0.2$ and the Reynolds number is given by $Re=1.0$. By varying the electrical properties of the middle layer, we demonstrate two regimes: attractive and repulsive behaviors. All phases share equal density and viscosity  $(\rho_1=\rho_2=$ $\rho_3,\mu_1=\mu_2=\mu_3)$.
\begin{figure}[htbp]
	\centering
	\includegraphics[width=0.27\textwidth]{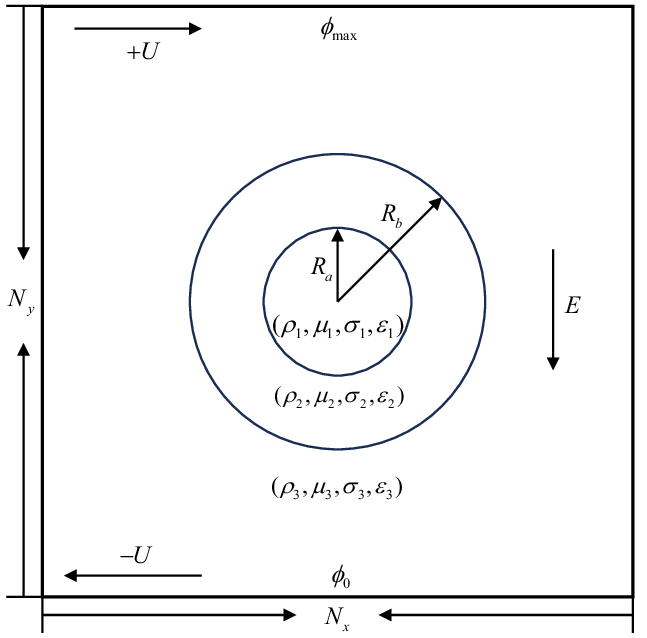}
	\caption{Schematic of a compound droplet in a confined shear flow..}
	\label{fig12}
\end{figure}

Fig. \ref{fig10} illustrates the interaction dynamics for a coalescence regime , where the intermediate layer properties are set to $S_{23} = 8.0, R_{23} = 6.0$ and $S_{12} = 1.0/8.0, R_{12} = 1.0/6.0$. Because the permittivity ratio exceeds the conductivity ratio ($S_{23} > R_{23}$), the tangential Maxwell stresses along the outer interface induce a robust electrohydrodynamic  flow directed from the poles towards the equator. This pole-to-equator flow continuously sweeps the continuous-phase fluid away from the narrow gap between the two droplets,  accelerating the film drainage process. Coupled with the inherent electrostatic dipole-dipole attraction, the droplets approach each other. The simulation captures a distinct sequential coalescence: the outer shells rupture and merge first, followed by the secondary film drainage of the intermediate phase, which leads to the coalescence of the two inner cores within the newly coalesced outer droplet.

Conversely, Fig. \ref{fig11} depicts a repulsive regime with $S_{23} = 0.2,R_{23} = 1.04$ and $S_{12} = 5.0,R_{12} = 1.0/1.04$. Under the condition of $R_{23} > S_{23}$, the accumulation of free charges at the interface reverses the direction of the tangential electric stresses, thereby driving the EHD flow from the equator towards the poles. As the fluid is pumped towards the pole regions, it collides in the narrow gap between the droplets, generating a localized high-pressure zone and an outward hydrodynamic flow. This EHD-induced hydrodynamic repulsion is sufficiently strong to overcome the electrostatic dipole attraction. Consequently, instead of undergoing film thinning, the adjacent interfaces are pushed apart, and the compound droplets exhibit a stable separation behavior, moving progressively further away from each other as time advances.

\subsection{Deformation of a compound droplet under confined shear flow}
In this section, we investigate the deformation behavior of a compound droplet subjected to a confined shear flow. The baseline geometric parameters of the compound droplet are set as follows: inner droplet radius $R_a = 15.0$, outer droplet radius $R_b = 30.0$. Fig. \ref{fig12} delineates the schematic configuration of the considered problem. The Reynolds number and electrical Reynolds number are set to $Re = 0.1$ and $Re_E = 0.01$, respectively. The resulting deformation of the droplet is dependent on two key dimensionless parameters: the electric capillary number $Ca_E$ and the confinement ratio $W_c$.

\begin{figure}[htbp]
	\centering
	\subfigure{\includegraphics[width=0.26\textwidth]{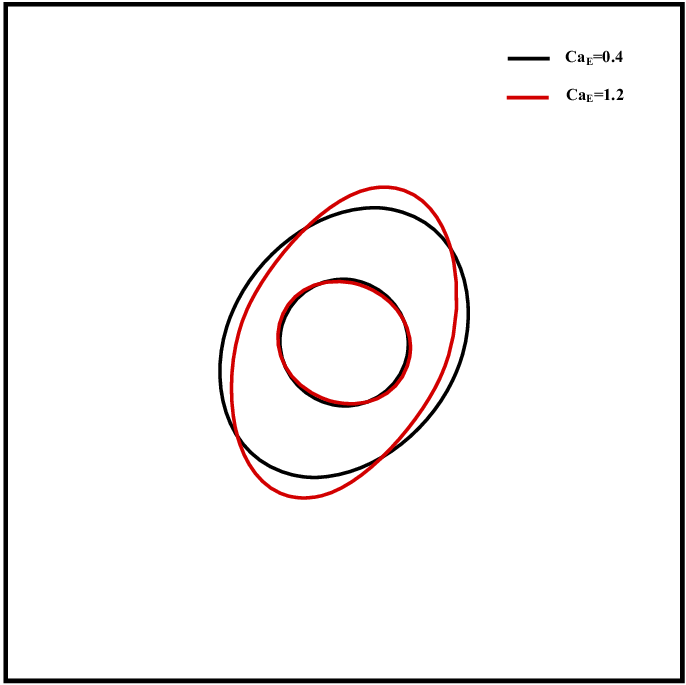}\label{fig13a}}\quad\quad
	\subfigure{\includegraphics[width=0.36\textwidth]{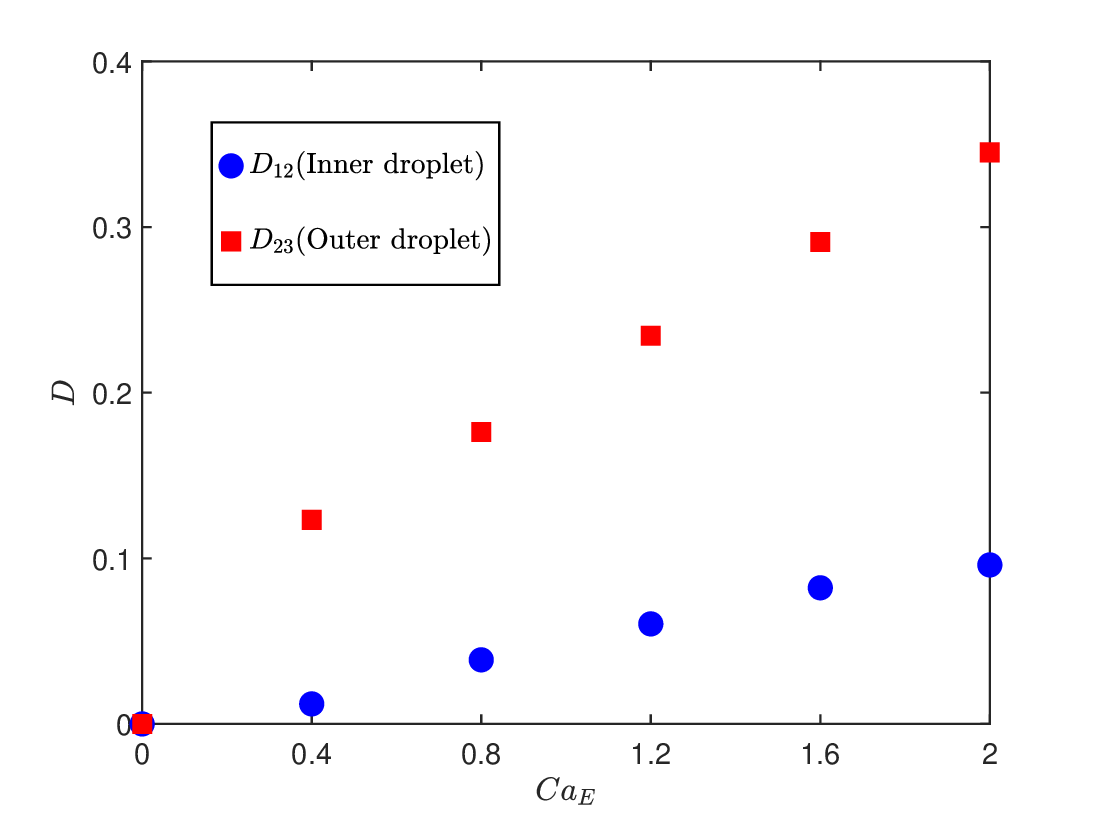}\label{fig13b}}
	\caption{(a) The compound droplet shapes for different values
		of $Ca_E$; (b) alteration of the steady-state deformation parameter with $Ca_E$.}
	\label{fig13}
\end{figure}

We first examine the influence of $Ca_E$ on droplet deformation under a constant  capillary number $Ca = 0.1$. For this case, the permittivity and conductivity ratios of the interfaces are specified as $S_{12} = 2.0$, $S_{23} = 0.5$, and $R_{12} = 0.5$, $R_{23} = 2.0$, respectively. Fig. \ref{fig13a} illustrates the relationship between $Ca_E$ (ranging from 0.4 to 2.0) and the deformation parameters of both the inner and outer droplets. As $Ca_E$ increases, the deformation parameters of both droplets increase monotonically, indicating a stronger stretching effect induced by the electric field. Notably, the deformation of the outer droplet is significantly more pronounced than that of the inner core. This is primarily because the outer droplet is subjected to the direct shearing effect of the external flow. Furthermore, the different electro-physical properties among the fluid phases produce a mismatch in both viscous and electrical stresses at the interfaces, which together drive this differential deformation.

\begin{figure}[htbp]
	\centering
	\subfigure[$W_c=0.3$]{\includegraphics[width=0.2\textwidth]{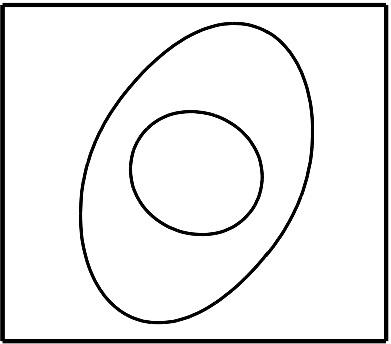}}\quad
	\subfigure[$W_c=0.5$]{\includegraphics[width=0.2\textwidth]{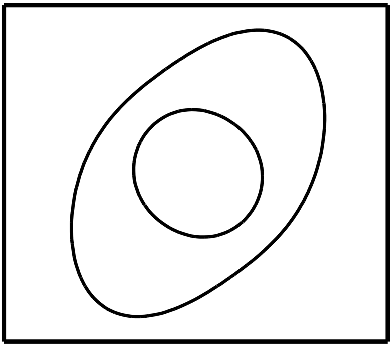}}\quad
	\subfigure[$W_c=0.6$]{\includegraphics[width=0.2\textwidth]{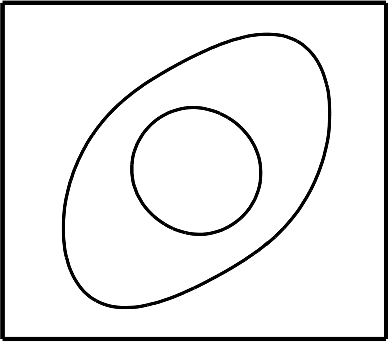}}\quad
	\subfigure[$W_c=0.75$]{\includegraphics[width=0.2\textwidth]{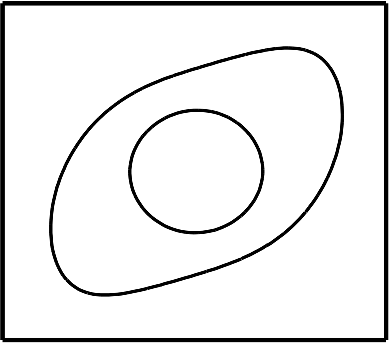}}
	\caption{Steady-state morphologies of the compound droplet at different confinement ratios $W_c$: (a) $W_c=0.3$, (b) $W_c=0.5$, (c) $W_c=0.6$, (d) $W_c=0.75$.}
	\label{fig14}
\end{figure}

The orientation of the compound droplet is also highly sensitive to variations in $Ca_E$. Figure \ref{fig13b} displays the steady-state snapshots of the compound droplet at $Ca_E = 0.4$ and $1.2$. The morphological comparison reveals that a higher $Ca_E$ leads to a larger inclination angle between the droplet's major axis and the horizontal direction. Physically, the droplet orientation is governed by the competition among the hydrodynamic shear stress, interfacial tension, and Maxwell stresses. As $Ca_E$ increases, the Maxwell stresses at the interface are significantly enhanced. This potent electrohydrodynamic force disrupts the shear-dominated stress balance, compelling the droplet to stretch intensely along the electric field axis. Consequently, the droplet resists the shear flow-induced alignment, resulting in an increased inclination angle relative to the horizontal streamlines.

Furthermore, we also examine how the confinement ratio $W_c$ affects the droplet inclination angle. For these cases,  $Ca_E=1.0$ and $W_c$ takes values $0.3, 0.5, 0.6, \text{and } 0.75$. The results in Fig. \ref{fig14} show an inverse relationship: a larger $W_c$ leads to a greater rightward tilt, meaning a smaller orientation angle with the horizontal axis. This behavior arises from competition between the vertical electric force and the horizontal hydrodynamic shear force. Stronger confinement amplifies the local hydrodynamic shear stresses. As $W_c$ increases, the horizontal shear gradually overcomes the vertical electrohydrodynamic stretching force that tries to keep the droplet upright, forcing the droplet to tilt further horizontally. 
\begin{figure}[htbp]
	\centering
	\includegraphics[width=0.4\textwidth]{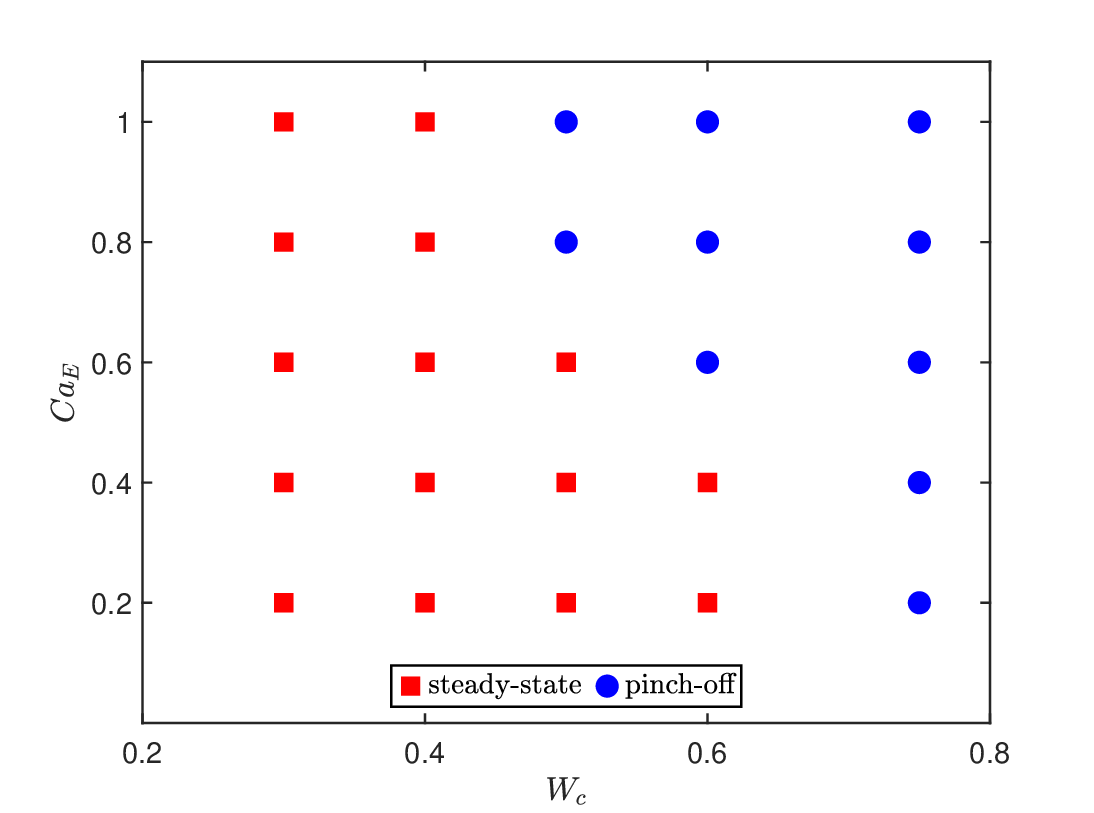}
	\caption{Regime  showing the droplet morphology for different values of $Ca_E$ and $W_c$.}
	\label{fig15}
\end{figure}

Finally, to delineate the stability of the compound droplet under the combined influence of the electric field and confinement, we present the regime diagram in the $Ca_E$-$W_c$ parameter space, as depicted in Fig. \ref{fig15}. For this case, the capillary number is fixed at $Ca = 0.5$, and the electrical property parameters are set to $S_{12} = 0.5$, $S_{23} = 2.0$, and $R_{12} = 2.0$, $R_{23} = 0.5$. This phase diagram covers a broad parameter range, with $Ca_E$ ranging from $0.2$ to $1.0$ and $W_c$ from $0.3$ to $0.75$. The steady-state deformation and  pinch-off regimes are distinctly denoted by red squares and blue circles, respectively. The evolution of the phase boundary indicates that the critical $Ca_E$ threshold triggering droplet pinch-off decreases significantly as the confinement effect intensifies.  In the weakly confined regime, the droplet can sustain stronger electric stretching, requiring a higher $Ca_E$ threshold to initiate pinch-off. However, as the degree of confinement increases, the severe spatial restriction forces the droplet into a highly unstable state. Under such strong confinement, the droplet becomes significantly more susceptible to rupture, meaning that a relatively lower $Ca_E$ is sufficient to overcome the interfacial tension and trigger  pinch-off.

\section{Conclusion}
\label{sec:conclusion}
In this study, a thermodynamically consistent phase-field model for three-phase EHD flows was proposed and numerically implemented using a mesoscopic LB method. Unlike conventional phenomenological approaches, the governing equations of the proposed model were strictly derived from the Onsager variational principle. This theoretical framework intrinsically ensures the thermodynamic consistency of the complex multiphase system and accurately incorporates the crucial surface charge convection mechanism without relying on a priori ad-hoc assumptions.

The accuracy, stability, and reliability of the developed LB method framework were rigorously verified through a comprehensive series of benchmark tests.  The excellent agreement between our simulations, analytical solutions, and existing literature confirms the model's robust capability in handling intricate multiphase interfaces and multiphysics coupling. The focus on the EHD behaviors of compound droplet coalescence, separation, and shearing not only validates the model but also reveals the highly nonlinear characteristics of three-phase systems. The results illustrate that the precise morphological evolution of compound droplet is intricately dictated by the interplay between electric field intensities, surface charge distributions, and external shear forces. This study ultimately provides a powerful, physics-preserving tool for revealing the underlying mechanisms of complex multiphase electrohydrodynamics.




\end{document}